%% file: main.tex
\documentclass[acmsmall]{acmart}

\usepackage{wrapfig}
\usepackage{fancybox}
\usepackage{multirow}
\usepackage{rotating}
\usepackage{makecell}
\usepackage{xcolor}
\usepackage{enumitem}
\usepackage[T1]{fontenc}
\usepackage[utf8]{inputenc}

\usepackage{amssymb}

\definecolor{myOrange}{RGB}{232,142,69}
\definecolor{myBlue}{RGB}{93,116,162}
\definecolor{myRed}{RGB}{182, 40, 43}

\newcommand{\calloutbox}[1]{%
  \begin{Sbox}
  \begin{minipage}{\dimexpr0.73\linewidth-3\fboxsep-3\fboxrule}
    #1
  \end{minipage}%
  \end{Sbox}
  \begin{center}
  \setlength{\fboxrule}{0pt} 
  \setlength{\fboxsep}{4pt} 
  \fcolorbox{black}{blue!6}{\TheSbox}
  \end{center}
}

\newcommand{\myshade}[1]{%
  \begin{Sbox}
    \mbox{#1}
  \end{Sbox}
  \setlength{\fboxrule}{0pt} 
  \setlength{\fboxsep}{0pt} 
  \fcolorbox{black}{blue!10}{\TheSbox}
}

\newcommand{\noshade}[1]{%
  \begin{Sbox}
    \mbox{#1}
  \end{Sbox}
  \setlength{\fboxrule}{0pt} 
  \setlength{\fboxsep}{0pt} 
  \fcolorbox{black}{blue!0}{\TheSbox}
}

\newcommand{\DataTransformationShade}[1]{%
  \begin{Sbox}
    \mbox{#1}
  \end{Sbox}
  \setlength{\fboxrule}{0pt} 
  \setlength{\fboxsep}{1.5pt} 
  \fcolorbox{black}{myOrange!75}{\TheSbox}
}

\newcommand{\VisualTransformationShade}[1]{%
  \begin{Sbox}
    \mbox{#1}
  \end{Sbox}
  \setlength{\fboxrule}{0pt} 
  \setlength{\fboxsep}{1.5pt} 
  \fcolorbox{green!20}{myBlue!75}{\TheSbox}
}

\newcommand{\sqshade}[1]{%
  \begingroup
  \setlength{\fboxrule}{0pt} 
  \setlength{\fboxsep}{1pt} 
  \fcolorbox{black}{myOrange!30}{#1}
  \endgroup
}

\newcommand{\highlightshade}[1]{%
  \begingroup
  \setlength{\fboxrule}{0pt} 
  \setlength{\fboxsep}{0pt} 
    \fcolorbox{black}{blue!10}{#1}%
  \endgroup
}

\newcommand{\ProcessControlShade}[1]{%
  \begin{Sbox}
    \mbox{#1}
  \end{Sbox}
  \setlength{\fboxrule}{0pt} 
  \setlength{\fboxsep}{1.5pt} 
  \fcolorbox{black}{myRed!75}{\TheSbox}
}

\newcommand{\myhline}{\noalign{\global\arrayrulewidth=0.0001pt}\hline
                      \noalign{\global\arrayrulewidth=1pt}}

\definecolor{softblue}{rgb}{0.2, 0.4, 0.8} 
\newenvironment{edited}{\color{black}}{}
               
\AtBeginDocument{%
  \providecommand\BibTeX{{%
    \normalfont B\kern-0.5em{\scshape i\kern-0.25em b}\kern-0.8em\TeX}}}

\setcopyright{acmcopyright}
\copyrightyear{2024} 
\acmYear{2024} 
\setcopyright{acmlicensed}\acmConference[CHI '24]{Proceedings of the CHI Conference on Human Factors in Computing Systems}{May 11--16, 2024}{Honolulu, HI, USA}
\acmBooktitle{Proceedings of the CHI Conference on Human Factors in Computing Systems (CHI '24), May 11--16, 2024, Honolulu, HI, USA}
\acmDOI{10.1145/3613904.3642740}
\acmISBN{979-8-4007-0330-0/24/05}

\acmSubmissionID{3722}

\begin{document}

\title[Data Cubes in Hand]{Data Cubes in Hand: A Design Space of Tangible Cubes for Visualizing 3D Spatio-Temporal Data in Mixed Reality}


\author{Shuqi He}
\affiliation{%
  \institution{Xi'an Jiaotong-Liverpool University}
  \city{Suzhou}
  \country{China}
}
\orcid{0009-0002-6365-8806}

\author{Haonan Yao}
\affiliation{%
  \institution{Xi'an Jiaotong-Liverpool University}
  \city{Suzhou}
  \country{China}
}
\orcid{0009-0003-1735-6281}

\author{Luyan Jiang}
\affiliation{%
  \institution{Xi'an Jiaotong-Liverpool University}
  \city{Suzhou}
  \country{China}
}
\orcid{0009-0004-0744-4824}

\author{Kaiwen Li}
\affiliation{%
  \institution{Xi'an Jiaotong-Liverpool University}
  \city{Suzhou}
  \country{China}
}
\orcid{0009-0003-2064-6992}

\author{Nan Xiang}
\affiliation{%
  \institution{Xi'an Jiaotong-Liverpool University}
  \city{Suzhou}
  \country{China}
}
\orcid{0000-0003-4028-2287}

\author{Yue Li}
\affiliation{%
  \institution{Xi'an Jiaotong-Liverpool University}
  \city{Suzhou}
  \country{China}
}
\orcid{0000-0003-3728-218X}

\author{Hai-Ning Liang}
\affiliation{%
  \institution{Xi'an Jiaotong-Liverpool University}
  \city{Suzhou}
  \country{China}
}
\orcid{0000-0003-3600-8955}

\author{Lingyun Yu}
\authornote{Corresponding author}
\affiliation{%
  \institution{Xi'an Jiaotong-Liverpool University}
  \city{Suzhou}
  \country{China}
}
\orcid{0000-0002-3152-2587}

\renewcommand{\shortauthors}{He et al.}


\begin{abstract}
Tangible interfaces in mixed reality (MR) environments allow for intuitive data interactions. Tangible cubes, with their rich interaction affordances, high maneuverability, and stable structure, are particularly well-suited for exploring multi-dimensional data types. However, the design potential of these cubes is underexplored. This study introduces a design space for tangible cubes in MR, focusing on interaction space, visualization space, sizes, and multiplicity. Using spatio-temporal data, we explored the interaction affordances of these cubes in a workshop (N=24). We identified unique interactions like rotating, tapping, and stacking, which are linked to augmented reality (AR) visualization commands. Integrating user-identified interactions, we created a design space for tangible-cube interactions and visualization. A prototype visualizing global health spending with small cubes was developed and evaluated, supporting both individual and combined cube manipulation. This research enhances our grasp of tangible interaction in MR, offering insights for future design and application in diverse data contexts.
\end{abstract}

\ccsdesc[500]{Human-centered computing~Visualization design and evaluation methods}
\ccsdesc[500]{Human-centered computing~Mixed / augmented reality}
\keywords{tangible interaction, spatio-temporal data, mixed reality}

\begin{teaserfigure}
 \centering
 \includegraphics[width=\textwidth]{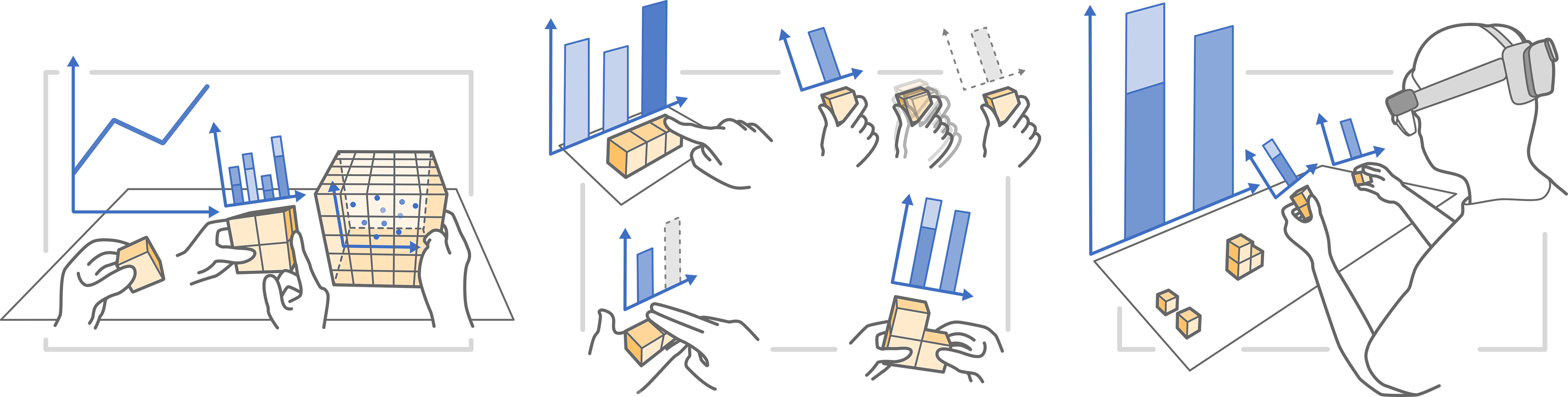}
 \caption{We explore design considerations on the size, multiplicity, interaction and visualization mappings of tangible cubes in MR environments. Left: the \textit{Size-Multiplicity Dynamics} showcase how the creation of combinations and varying cube sizes can dictate distinct visualization and interaction paradigms. Middle: the \textit{Interaction-Visualization Mappings} reveal how tangible actions are transformed into visual outcomes. Right: the \textit{Integrated MR System} synthesizes these design considerations into a cohesive tangible cube system, illustrating a user engaging with data visualizations in an MR environment.}
 \Description{An image divided into three sections. The left section displays three cube sizes, with medium and large cubes composed of smaller ones, representing different visualization and interaction styles. The middle section illustrates icons of four interactions and their corresponding visual outcomes. The right section depicts a user wearing a head-mounted display, holding cubes, with data visualizations both on the cubes and anchored to a tabletop.}
 \label{fig:teaser}
\end{teaserfigure}


\maketitle

\section{Introduction}

The recent advancement of immersive technologies has spawned innovative approaches to data interaction and visualization. Among these, mixed reality (MR) environments emerge as a prominent immersive modality. The term MR refers to a continuum that integrates and fosters interaction between the real and virtual environments, as defined by Milgram et al. in the early 1990s \cite{milgram1995augmented, speicher2019mixed}. By bridging the gap between the real and virtual realms, MR can enhance data exploration experiences by providing more immersive and adaptable solutions \cite{Lee22ADesignSpace}. Notably, various tangible user interfaces have been implemented as data embodiments to represent virtual data within the real world. These tangibles can be active proxies that are physically manipulated by the users for representing abstract data \cite{Satriadi23Active}. As tangibles are inherently part of the real environment, they offer a practical means for users to engage physically and seamlessly with abstract data. Consequently, tangibles serve as valuable tools for optimizing data tasks in MR \cite{feick2020tangi}.


\vspace{2pt}
Tangible designs encompass a range of geometries, such as globes \cite{Satriadi22Globes}, cylinders \cite{Jofre15} and customized scale models that mimic real-wrold structures \cite{ens2020uplift, Hull22SimulWorlds, Satriadi23Active}. Among these diverse shapes, cubes have emerged as a particularly prominent form, attracting considerable research attention due to their unique affordances and versatile applications \cite{fitzmaurice1995bricks, pla2013display}. Focusing on a single geometry, such as cubes, allows for a more systematic exploration. This approach enables an in-depth analysis of the ways people handle and interact with cubes as a particular tangible object, examining its unique affordances. This helps avoid the complexity and potential confusion that may arise from juggling too many parameters associated with multiple shapes.

\vspace{2pt}
The advantages of employing cubes as tangible interfaces can be summarized across several distinct aspects. First, similar to other tangible interfaces, their three-dimensional (3D) structure facilitates the manipulation of 3D data through direct interaction. Second, cubes have established a natural mental model of interaction, owing to the numerous cube-based designs already embedded in everyday life. For instance, the game mechanics of the renowned Rubik's cube exemplify cube-enabled tangible interactions \cite{bergig2011out}. As users have experience manipulating commonplace cubic objects and toys such as the Rubik's cube, they are well-acquainted with the underlying tangible concept and the fundamental interactions it allows.
While their 3D structure and established mental model are common to many tangible geometries, it is the cube's modular scalability that sets them apart. Unlike other tangible geometries such as globes or pyramids, the simple architecture and modular nature of cubes allow them to be easily and stably combined in all three axes to scale up to varied constructive assemblies without altering their underlying structure. As a result, the planar surfaces and the enclosed volume can be designed to accommodate different interaction styles and display modalities \cite{Hornecker2006GettingAG}. The ease of manipulation, intuitive interaction and unique scalability afforded by tangible cubes are especially valuable in the context of multi-dimensional data, the complexity of which may otherwise be challenging to decode with traditional visualization techniques. 

\vspace{2pt}
Multi-dimensional data contains rich information and intricate relationships between variables \cite{pastizzo2002multidimensional}. However, analyzing multi-dimensional data can be challenging due to the high number of dimensions involved. To understand this type of data, various aspects of visual representations can be altered, including the ordering, spatial arrangement, or the use of different visual channels. Moreover, multiple interaction operations, such as sorting, filtering and selection, need to be performed and switched between to effectively navigate and explore the data. 
Spatio-temporal data, a form of multi-dimensional data, encompasses information collected across both spatial and temporal dimensions. It has found applications in diverse research domains, where it is collected, analyzed, and visualized. Examples include depicting historical evolution \cite{deng2009spatio}, examining motion trajectories \cite{demvsar2010space, kang2010mining, gudmundsson2017spatio} and predicting natural disasters \cite{han2022data, aubrecht2013spatio, li2016spatio}. The focus on spatio-temporal data is driven by two motivations. First, it has real-life relevance. Despite the technical-sounding name, spatio-temporal data is familiar to many in applications like weather forecasting and GPS navigation. Second, spatio-temporal data is compatible with MR for effectively presenting its complexities, which are not as readily communicated through simpler datasets or traditional 2D displays. This choice leverages MR's full potential, consistent with its practical uses in current research.

\vspace{2pt}
The concept of the \textit{space-time cube} is developed to effectively represent spatio-temporal data within a 3D cube. Originally intended to analyze socio-behavioural patterns of individuals over time \cite{bach2014review, ilagcrstrand1970people}, the concept is now employed in numerous applications for visualizing data \cite{gatalsky2004interactive, fisher2005visualization, li2010visual}. Empirical evidence has demonstrated that it facilitates a more efficient understanding of complex spatio-temporal patterns \cite{kristensson2008evaluation}. 
More recently, the space-time-cube concept has been applied in tangible systems for visualizing and comparing temporal trends in complex data such as energy consumption for different buildings \cite{ens2020uplift}.
Given the effectiveness of the space-time cube concept, it is imperative to propose using cubic tangibles as a compelling means to represent this type of data. 

\vspace{2pt}
Despite growing interest in tangible cubes and their application in MR environments, a notable research gap exists regarding how tangible cubes can be applied for visualizing and interacting with multi-dimensional data such as that represented by the space-time cube. The absence of a unifying framework outlining possible interactions and visualization commands limits our ability to explore the full range of design possibilities offered by tangible cubes. 

\vspace{2pt}
Therefore, this work aims to address this research gap by proposing a design space of tangible cubes for visualizing 3D spatio-temporal data in MR. We begin by evaluating related work employing tangible cubes of varying sizes and multiplicity, and summarizing its implications on interaction and visualization spaces. Next, we introduce the design space mapping user interaction actions to visualization commands. This mapping was created based on a workshop of $24$ participants brainstorming interaction pairings under $10$ different data tasks. We demonstrated the utility of our design space by creating a prototype adopting the most intuitive interactions identified. The prototype supports manipulation of individual cubes, as well as the combination of individual cubes into assembled structures. 

\vspace{0.5em}
Our work contributes to the field of mixed reality interactions and visualizations through several key aspects. First, we identify design opportunities and constraints associated with the \textbf{size and multiplicity} of tangible cubes, which can shape diverse interaction techniques and visualization strategies, providing a basis for future exploration in MR environments. Second, we introduce a user-inspired \textbf{design space} of tangible cubes that offers intuitive mappings between interactions and specific visualization commands in MR. Furthermore, our \textbf{proof-of-concept prototype} showcases the practicality of using tangible cubes for visualizing spatio-temporal data, such as global health spending, in real-world scenarios. Overall, this study advances our understanding of tangible-cube-enabled interactions and visualizations in MR environments, providing valuable insights for future development and implementation.

\section{Related Works}
\begin{edited}
The concept of tangible user interfaces (TUIs) has been integral to the evolution of interaction design in immersive environments. This section first provides an overview of related works on TUIs in general, then narrows down to discuss the specialized application of tangible cubes, detailing their contributions to interaction and visualization design in MR.
\end{edited}

\begin{edited}
\subsection{Tangible User Interfaces}
TUIs represent a shift in interaction by integrating graspable physical objects into the digital interface to take advantage of the user's familiarity with interacting with everyday objects. Early TUI designs such as \textit{Tangible Bits} \cite{Ishii97TangibleBits} and \textit{Urp} \cite{Underkoffler99Urp} showcased how physical models such as graspable objects, interactive surfaces and ambient displays can be projected with shadows, reflections and other graphical information for intuitive analysis of different conditions. The historical and conceptual foundations of TUIs have been comprehensively studied, tracing their evolution and their potential to transform digital interactions across various domains \cite{Shaer10TUIPastPresentFuture}.

\vspace{2pt}
In their comprehensive survey on spatial interfaces, Besan\c{c}on et al. \cite{Besancon:2021:SAS} acknowledged the unique role of TUIs as a natural and flexible means for visualizing 3D data, summarizing their application in tasks like volumetric view and object manipulation, manipulation of visualization widgets, along with data selection and annotation. Notably, tangible interaction was found to be used the most for manipulating 3D widgets when compared with other interaction paradigms, attributed to their intuitive handling and natural positioning in 3D space \cite{Besancon:2021:SAS}.

\vspace{2pt}
Recent work has further explored how tangibles can be utilized in visual analytics. For instance, Ens et al. presented \textit{Uplift} \cite{ens2020uplift}, combining a tabletop display with tangible widgets in augmented reality for collaborative tasks in building energy analysis. Users place physical building models on an interactive table to view corresponding energy usage data. The tangibles serve as physical referents to tie the abstract data to real-world artifacts. Hull et al.'s work \cite{Hull22SimulWorlds} focuses on using physical scale models to overlay, compare and integrate multiple datasets to provide context. Satriadi et al. introduced the concept of Active Proxy Dashboard \cite{Satriadi23Active} where users manipulate tangible scale models in mid-air to filter and query data visualizations on a display.

\end{edited}

\subsection{Tangible Cube Interfaces}

\begin{edited}
    Tangible cubes are a unique subset of TUIs with modular scalability that allows them to be combined in multiplicity. In related works, cubic interfaces have been applied across various domains and design scenarios. To gain a comprehensive understanding of tangible cubic interfaces in immersive environments, we employed a targeted literature review strategy focusing specifically on designs involving cubic tangibles in MR. We employed keywords such as ``augmented reality'', ``mixed reality'', ``tangible cubes'', and ``visualization''. The search encompassed conferences and journals in several dimensions, including Human-Computer Interaction (CHI, INTERACT), Virtual and Augmented Reality (ISMAR, 3DUI, SIGGRAPH), Visualization (EGVE, PacificVis) and Tangible Interaction (TEI, ITS). Priority was given to papers published in the last 15 years.
\end{edited}
Our analysis centered on the range of interactions these systems support and the immediate visual outcomes resulting from these interactions. Furthermore, we examined taxonomies and frameworks related to space-time cube visualizations and tangible cube interactions. In the following sections, we present a review of relevant works in the areas of tangible cube interfaces, space-time cube visualization taxonomies, and tangible cube interaction frameworks.

\vspace{2pt}
The application of tangible cubes in MR has been investigated in interaction design for educational purposes. Juan et al. \cite{juan2010tangible} explored using three small tangible cubes side by side for educating children about endangered animals. Each side of the cube features distinct identification markers, which present different information and videos of virtual animals on the cube surfaces upon rotation. More recently, Olim et al. \cite{olim2020augmented} employed five tangible cubes for learning chemical elements from the periodic table, with each cube representing a single chemical element. Each facet of the cube contains information and facts about the element to be prompted upon rotation. Similarly, Song et al. \cite{song2019turtlego} utilised AR tangible cubes as teaching aids to enhance children's spatial abilities. By stacking and assembling cubes in various orientations, users interact with a virtual turtle moving atop the cubes, gaining spatial awareness through this gamified experience. 

\vspace{2pt}
Tangible cubes have also been implemented for 3D objects or volume manipulation. They can serve as direct carriers of virtual objects, as demonstrated by Issartel et al. \cite{issartel2016tangible}, who proposed a cubic tangible volume concept for grasping and manipulating virtual objects in an MR 3D environment. Volume selection and grasping are initiated by pressing and applying finger pressure above a certain threshold. The selected 3D volume can then be manipulated and relocated through rotation and translation of the tangible cube to various orientations and locations within the environment. Alternatively, tangible cubes can function as control mediums, with their effects mapped to the objects being controlled. For instance, the \textit{CubTile} design \cite{de2008cubtile} employs a translucent cube for manipulating 3D objects projected on a large display environment, supporting selection, translation, rotation, and scaling.

\vspace{2pt}
Of particular interest is the use of tangible cubes for data interaction. Sifteo cubes \cite{merrill2012sifteo}, tangible cuboids developed by the MIT media lab, have been employed as touchscreen interfaces for data queries. Langner et al. \cite{langner2014cubequery} created a musical dataset query system based on Sifteo cubes. In their design, each cube represents an individual data search parameter. By changing the arrangement of the Sifteo cubes on a table surface, search parameters can be combined to filter the database. The search results are then displayed on an interactive tabletop surface in a grid format.

\subsection{Space-Time Cube Visualization Taxonomies}
Space-time cube visualizations have been widely adopted for their intuitive representation of complex data across spatial and temporal dimensions within a sleek 3D volume. For instance, Zhang et al. \cite{Zhang:2022:TimeTables} developed a prototype system in virtual reality to facilitate data exploration using space-time cubes. Despite their effectiveness, the terminology and visualization commands associated with space-time cubes can be ambiguous. In an effort to establish consistent languages and a unified framework, several taxonomies have been proposed to characterize the visualization commands specifically tailored for space-time cubes. For instance, Bach et al. devised a theoretical taxonomy detailing the possible operations that can be performed on a generalized space-time cube visualization \cite{bach2014review, bach2017descriptive}. They offered a detailed classification of the operations, focusing on the \textit{extraction}, \textit{flattening}, \textit{geometry transformation} and \textit{content transformation} aspects of the visualization. Bach's taxonomy is highly pertinent to our framework and serves as a foundational reference in guiding the brainstorming and development of visualization commands, informing researchers about the most effective interaction techniques for specific data tasks and contexts. Although Bach's work remains theoretical and descriptive in nature, we aim to build upon it by refining and extending the framework to reveal intuitive visual mappings enabled by user-defined actions.

\subsection{Tangible Cube Interaction Frameworks}
Several design space works have investigated interaction techniques for tangible cubes, with researchers exploring various dimensions of gestural design, interaction and visualization. Valdes et al. \cite{valdes2014exploring} explored data query manipulations using gestural interactions based on Sifteo cubes. They adopted a task-driven approach in their user-elicitation study, where participants are asked to propose \textit{gestures} for manipulating the Sifteo cubes in completing a specific task. The resulting taxonomy classifies the user-defined gestures based on their interaction space (on-surface, on-bezel, in-air), interaction flow (continuous, discrete), and number of hands and tangibles involved (cardinality). More closely related to our scope of immersive spatio-temporal visualization, Cordeil et al. \cite{cordeil2017design} proposed a design space discussing tangible design considerations from several broad aspects, including the size of the interaction space, degree of physicality, specific interaction support (navigation and menu) and the display space. They then used this descriptive design space for generating three design examples, one of which is a touch-sensitive cube allowing volumetric selection of data through surface gestures. Most recently, Potts et al. introduced \textit{TangibleTouch} \cite{potts2022tangibletouch}, a toolkit for designers to prototype and evaluate gestural interactions. As part of their design process, they proposed a design space specifically for \textit{surface-based gestures} on tangible cubes, covering interactions such as tap, pinch, swipe, path and cover. The gesture inputs can be mapped to MR output spaces. 

\vspace{1em}
Collectively, these works contribute to various aspects of tangible cube design while also highlighting the need for continued exploration. Valdes et al.'s work \cite{valdes2014exploring} did not specifically focus on immersive modalities, which may restrict the applicability of their findings to MR environments. Cordeil et al.'s work \cite{cordeil2017design} discussed tangibles in a general sense, rather than specifically concentrating on tangible cubes or proposing specific visual mappings. Potts et al.'s work \cite{potts2022tangibletouch} targeted only surface-gesture-based interactions without exploring the full range of user action possibilities. Therefore, in this work, we aim to propose a more focused and complete design space investigating the interaction opportunities offered by tangible cubes in MR environments.

\section{Consolidating Interaction Taxonomy}
\label{sec:taxonomy}
The concept of tangible cubes offers a wide range of design possibilities. Based on our review of prior works, we have identified several key aspects of design choices that fundamentally determine the affordances of tangible cubes. These design dimensions encompass the \textbf{\textit{size}}, \textbf{\textit{interaction space}}, \textbf{\textit{visualization space}} and \textbf{\textit{multiplicity}} of the tangible cubes. The surveyed related works are summarized in \autoref{tab:summary}. In the following sections, we discuss the implications of cube size for the opportunities and constraints in interaction and visualization, as well as the ways multiple cubes can be used in conjunction for data manipulation.

\input{table/cube_taxonomy}

\subsection{Cubes of Varied Sizes}

\begin{figure}
  \centering
  \includegraphics[width=0.5\linewidth]{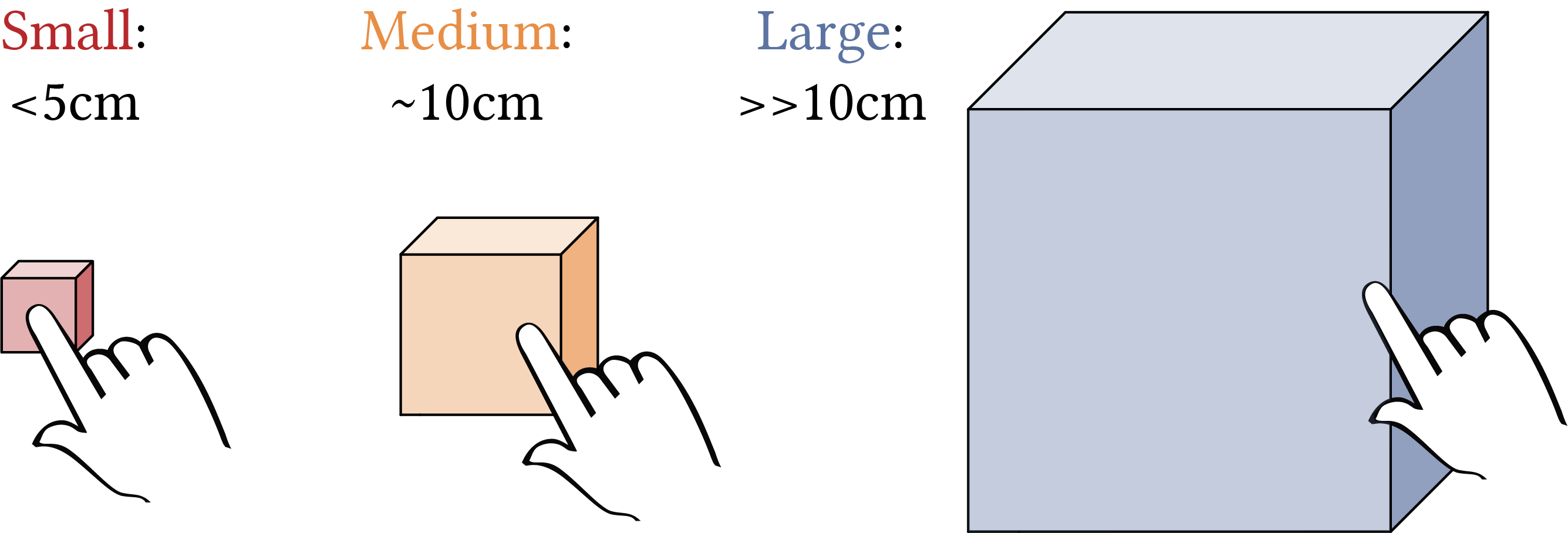}
  \caption{Tangible cubes of \textit{small}, \textit{medium} and \textit{large} sizes.}
  \Description{A small red cube, a medium yellow cube and a large blue cube, each with a finger pointing at them for size reference. Above each cube is a text label that suggests the size of each cube size category: small <5cm, medium ~10cm, large >>10cm.}
  \label{fig:Cube_Size.png}
\end{figure}

Three distinct dimensions have emerged in our analysis of tangible cube sizes (\autoref{fig:Cube_Size.png}). \textit{Small} cubes, with an edge length of approximately $5$cm or less, are designed for easy handling and precision grip, facilitating fine motor manipulation. They are typically meant to be picked up and oriented, making them ideal for educational \cite{juan2008augmented, zhou2004magic, olim2020augmented, lee2020using, song2019turtlego} or entertainment applications \cite{bergig2011out, grandhi2019playgami}. 
\textit{Medium}-sized cubes, comparable in size to a conventional Rubik's cube with edge lengths around $10$cm, can still be held with one hand. However, manipulation usually requires the support of the other hand, allowing users to pinpoint and orient areas of interest. 
\textit{Large} cubes, with considerably longer edges and heavier weights, are not designed to be picked up freely \cite{de2008cubtile}. Instead, they are typically stationary on a surface and primarily function as interactive displays. Users can physically move around the cubes, adjusting their viewing angles to find the optimal position of observation. Sometimes, designs can utilize a mix of these \textit{various} sizes within the same setup for a mixture of tactile experiences, offering different levels of motor controls and visualization possibilities \cite{cleto2020Code,hsu2022based}.
\vspace{5pt}

\subsection{Interaction Space}

The term ``interaction space'' encompasses the ways in which tangible cubes are manipulated and adjusted to achieve specific outcomes (\autoref{tab:summary}). \textit{Orientation}, for instance, involves altering the cube's 3D position by adjusting its rotation and tilt angles. \textit{Translation} refers to altering the cube's horizontal or vertical position without changing its orientation. \textit{Combination} denotes the act of positioning a tangible cube alongside another cube or object. \textit{Surface} interactions involve specific manipulations achieved through gestures or trajectories. \textit{Transformation} refers to structural changes or reshaping of the cube itself, including actions such as folding \cite{grandhi2019playgami}, shuffling \cite{bergig2011out}, and spinning \cite{lee2010tangible}. These transformations are a unique category of interactions and are dependent on the cube's specific material and composition.

\subsection{Visualization Space}
The term ``visualization space'' refers to ways visualizations are presented and organized in relation to the tangible cubes. We have identified several dimensions of visualization space in related literature, including \textit{Overlay}, \textit{Above}, \textit{Side}, \textit{Display}, \textit{Inside} and \textit{Around}.

\vspace{2pt}
\begin{description}[nosep,leftmargin=1.5em,labelindent=0em,leftmargin=!,labelindent=!,itemindent=!,font=\normalfont\itshape]%
\item[\textbf{Overlay}:] The \textit{overlay} method maps visual information directly onto the surface of the cube, rendering sides of the cubes as display media. 

\vspace{2pt}
\item[\textbf{Above}:] In contrast, \textit{Above} visualizations extend beyond the 2D cube surfaces. They use the cube surface as a fixation point, but allow the visualization to be projected into the 3D space directly above a cube surface. This approach allows users to manipulate the cube without physically obstructing the visual information. 

\vspace{2pt}
\item[\textbf{Side}:] \textit{Side} visualizations are positioned in open spaces adjacent to the tangible cube. This technique allows the user to view the visual representations from multiple perspectives, hence creating an immersive interaction experience.

\vspace{2pt}
\item[\textbf{Display}:] \textit{Display} visualizations make use of separate display media, such as a workstation, interactive tabletop, or touchscreen. This technique decouples the physical manipulation of the cube from the visualization itself, offering more flexibility in interaction scenarios.

\vspace{2pt}
\item[\textbf{Inside}:] The \textit{Inside} visualization technique renders the visual information within the cube, where the user can explore the visualization by looking into or through the cube. This approach of ``containing'' the visual information has been utilized as an intuitive way for selecting and transporting visual objects in an immersive environment \cite{issartel2016tangible}.

\vspace{2pt}
\item[\textbf{Around}:] Lastly, the \textit{Around} technique involves projecting information laterally around the cube, creating a visual space that encompasses the entire cube and potentially offering a 360-degree view. 
\end{description}

\subsection{Multiplicity}
Multiplicity refers to the quantity of tangible cubes utilized in a design scenario, employing a single or multiple cubes.

\vspace{2pt}
\begin{description}[nosep,leftmargin=1.5em,labelindent=0em,leftmargin=!,labelindent=!,itemindent=!,font=\normalfont\itshape]%
\item[\textit{\textbf{Single}}]cube designs tend to focus on the interaction and data visualization on or around the cube. In these approaches, the cube often serves as a standalone device to enable a set of specific interactions, such as orientation, translation, surface interaction and transformation. These designs benefit from the simplicity of a single tangible cube and achieve visualization or object manipulation through adjusting the cube.

\vspace{2pt}
\item[\textit{\textbf{Multiple}}]cubes or a larger cube composed of multiple smaller cubes, on the other hand, have been utilized by related works to enhance functionality \cite{juan2010tangible,gong2019grey,ma2020mixed,langner2014cubequery,song2019turtlego,bergig2011out, lee2011two, lee2010tangible, cleto2020Code,hsu2022based}. However, few take advantage of their capacity to facilitate complex dataset manipulation. The use of multiple cubes can create ensembles of data that leverage the relationships between cubes through spatial arrangements. The multiplicity of the cubes can potentially enable complex, collaborative and scalable interactions.

\end{description}

\subsection{Identified Opportunities and Constraints}

Our review of previous research on tangible cube-based systems revealed several distinctive design patterns. These patterns offer insights into the interplay between the selection of cube size, multiplicity and the associated interaction and visualization spaces.

\vspace{2pt}
\begin{description}[nosep,leftmargin=1.5em,labelindent=0em,leftmargin=!,labelindent=!,itemindent=!,font=\normalfont\itshape]%
\item[\textbf{Size and Multiplicity:}] Small and medium-sized cubes can be used in multiplicity scenarios. This is particularly promising with small-sized tangible cubes, characterized by their portability, manageability and ease of manipulation. Large cubes, on the other hand, tend to be used in singularity, as relegated by their weight and size constraints. It's worth noting that even though Rinott et al.'s work \cite{rinott2013cubes} provides opportunities for inter-connections between multiple large cubes, single cubes are largely regarded as complete experiences with one-to-one input-output mappings. We therefore categorized it as \textit{Single} for multiplicity.

\vspace{2pt}
\item[\textbf{Size and Interaction Space:}] The size of a cube inherently influences the choice of the optimal interaction space. Small and medium-sized cubes are often paired with \textit{orientation} interactions, highlighting their suitability for tasks demanding precise hand movements and fine motor control. On the other hand, the larger surface area and immobile nature of the large cubes make them ideal for \textit{surface}-based interactions. This dynamic presents an inherent trade-off between ease of manipulation and the available surface area based on the cube's size.

\vspace{2pt}
\item[\textbf{Size and Visualization Space:}] The visualization space is also influenced by the size of the tangible cubes. For instance, medium-sized cubes often employ the \textit{overlay} visualization strategy, taking advantage of their adequate surface area that can accommodate the overlays without overcrowding the physical form. In contrast, the constrained surface area of small cubes makes them less suitable for \textit{overlay} visualizations, as their limited surface area may not offer sufficient space for effective visual feedback.

\hspace*{1em}
Both medium and large cubes are observed to use the \textit{inside} approach. Their larger internal volumes provide more expansive visual canvases that allow the inclusion of intricate visual elements. Again, this option is not used by small cubes likely due to the limited internal volume which may not effectively house such visualizations.

\hspace*{1em}
Remarkably, medium-sized cubes demonstrate the greatest versatility as they span a wide range of design choices in the visualization space. This flexibility stems from their balanced size, which provides ample surface area and inner volume while maintaining manageability and ease of manipulation when designed carefully.
\end{description}

\vspace{5pt}

\noindent
Altogether, our review and analysis of previous works illuminate the impact of tangible cube size on multiplicity, interaction and visualization spaces. These interdependencies manifest as careful design decisions, each of which can significantly influence the cube's affordances and applications. However, it is important to acknowledge that our findings predominantly originate from designs focusing on small and medium-sized cubes, with large-sized cubes being notably underrepresented in the literature. This distribution reflects the existing state of tangible applications and resonates with the inherent characteristics and constraints of tangible interfaces. Yet, it also creates an imbalance in our sample of works, which may place limitations on the inferences. Nevertheless, these insights shed light on the interplay between various design choices and contribute to defining the diverse design space of tangible cubes.

\section{User-Inspired Interaction Mapping}
 
Building upon the examination of existing tangible cube theories, we explore the relationship between the interaction space and visualization space. Consistent with the definitions outlined in \autoref{sec:taxonomy}, an interaction action denotes the user's physical manipulation of the tangible cubes. These actions can include operations such as \textit{orientation}, \textit{translation} or \textit{combination} of tangible cubes. Conversely, a visualization command refers to the initiation or modification of the visual information displayed in the visualization space. These commands can encompass operations such as \textit{showing}, \textit{flattening} or \textit{recoloring} a visualization of a space-time cube. For instance, the action \textit{cover} might be mapped to the visualization command \textit{hide}. In this context, the cascade $\textit{cover}\rightarrow\textit{hide}$ stipulates that when a user covers a tangible cube in the interaction space, the corresponding data is concealed in the resulting visualization. To explore further mappings between interaction actions and visualization commands, we conducted an ideation workshop. The ultimate goal was to derive a versatile, expressive and effective design space that is both user-informed and richly informative.

\subsection{Initial Set of Interaction and Commands}

Building upon the foundation laid by established taxonomies outlined in related works, we assembled an initial set of interaction actions: \textit{rotate}, \textit{translate}, \textit{shake}, \textit{cover}, \textit{swipe}, \textit{pinch}, \textit{path}, \textit{tap}, \textit{hover}, \textit{neighbor}, \textit{stack} and \textit{assemble} \cite{potts2022tangibletouch, valdes2014exploring}. Similarly, a repertoire of spatio-temporal visualization commands was constructed, encompassing \textit{chopping}, \textit{flattening}, \textit{recoloring}, \textit{combination} and \textit{re-scale} \cite{bach2014review,bach2017descriptive}.
This initial set was provided to the workshop participants as a starting point for inspiration to explore how actions on the tangible cube can be mapped to various visualization commands for spatio-temporal data. 

\subsection{Ideation Workshop}

\subsubsection{Participants} 
We invited a cohort of $24$ participants, comprised of one lecturer, two PhD students, $19$ master's students and two undergraduate students. With backgrounds in visualization, computer science and human-computer interaction, the participants had relevant knowledge about interactive MR to generate diverse ideas within the limited time frame of the workshop session. This workshop was approved by the university ethics committee.

\subsubsection{Sample Data and Tasks}
\label{subsec: datacontext}
\begin{figure*}[t]
  \centering
  \includegraphics[width=\linewidth]{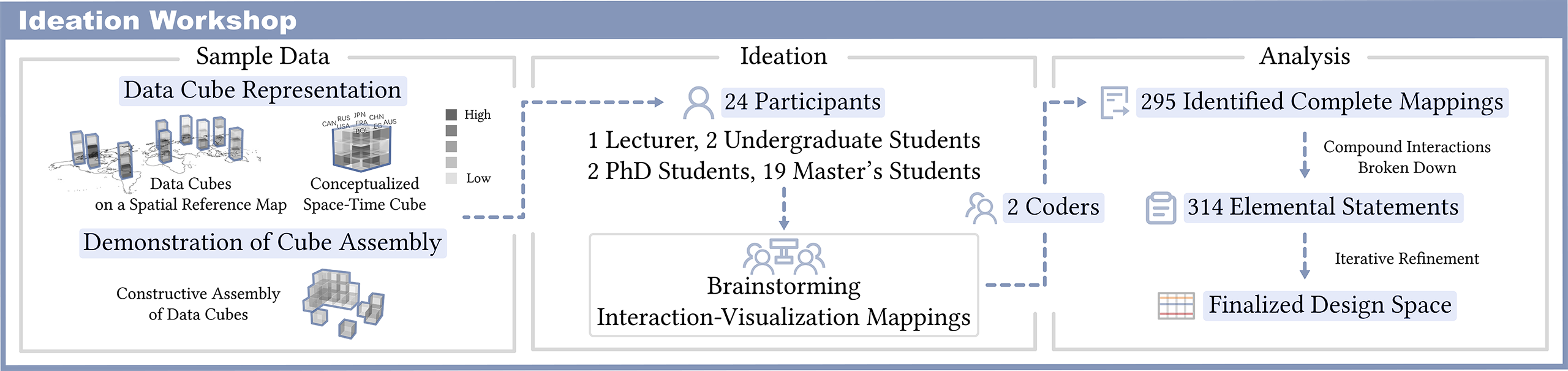}
  \caption{Flow chart of the ideation workshop and the subsequent analysis. Left: Illustration of the sample data, cube representation and assembly. Each column represents a different country, whereas the vertical axis represents time. These individual cubes can be manipulated to create constructive assemblies. Middle: Participants and process of the ideation workshop. Right: Analysis workflow decomposing compound interactions and coding elemental statements into the finalized design space.}
  \Description{A three-section flowchart illustrating the process of the ideation workshop. The left section displays the conceptual spatial arrangement of the case used in the workshop, this panel is divided into two sub-panels. The top panels shows stacks of cubes placed on a map, and a 3 by 3 by 3 cube formed like a rubik's cube representing the space-time cube. The bottom panel shows cubes stacked randomly forming a constructive assembly. The cubes are labelled as grey-scale heatmap style. The middle section shows the participants of the workshop, including 1 lecturer, 2 undergraduate students, 2 PhD students and 19 master’s students, arrows point to how they participated in the brainstorming using pictograms of people. The right panel is connected with the left by an arrow that says 2 coders. The follow chart continues with 295 identified complete mappings, which are broken down into 314 elemental statements and into the finalized design space through iterative refinement.}
  \vspace{-1em}
  \label{fig:workshop_stc}
\end{figure*}

To provide context and motivation for the guided brainstorming session, we utilized a global health expenditure dataset \cite{murray2023}. We extracted general government health expenditure data for nine countries, including Canada, USA, Japan, Bolivia, Russia, France, Egypt, China, and Australia. This data was visually represented through columns of cubes placed on a base map for spatial referencing, collectively forming a conceptualized $3\times3\times3$ space-time cube. Additionally, we showed that individual data cubes could be manipulated to create constructive assemblies, using physical cube props for demonstration. The left panel of \autoref{fig:workshop_stc} illustrates the data cube representations used in the workshop.

\vspace{2pt}
Each of the nine columns of the cube signifies health expenditure data from a distinct country. These data are segmented across three different periods (three stacked cubes in each column), each encapsulating a time scale of $10$ years. Therefore, each vertical stack of cubes corresponds to a $30$-year duration from 1990 to 2020. The topmost cube represents the most recent decade (2010-2020), the middle cube represents the previous decade (2000-2010), and the bottom cube represents the earliest decade (1990-2000). This conceptual model clarified the structure of the space-time cube and served to facilitate the brainstorming process.

\vspace{2pt}
To direct the brainstorming session, we proposed six distinct data tasks, including \textit{exploration}, \textit{combination}, \textit{difference}, \textit{snapshot generation}, \textit{annotation} and \textit{re-scale}. Each task provided one or more guiding questions (see \autoref{app: appendix1}) to provoke the participants to generate appropriate interaction-command pairs.

\subsubsection{Procedure}
The two-hour workshop started with the distribution of printed versions of the initial set of interactions and commands and idea collection sheets. The participants were then introduced to the research background and the space-time cube concept. We elaborated on the interaction and commands from the initial list by demonstrating individual operations with physical cube blocks of varied sizes. This ensured that the participants fully understood the distinctions between various interactions and commands. Following this, the task cases were presented one by one for brainstorming possible interaction-command pairs. These brainstormed ideas were then recorded by the participants on the data collection sheets which followed the structure: 

\vspace{1em}
\calloutbox{\begin{center}
``I would like to apply \underline{(interaction actions)} to achieve \underline{(visualization commands)}.''
\end{center}}
\vspace{1em}

We encouraged the participants to go beyond the initial set of interactions and commands and think outside the boundaries of the given tasks, providing further interaction-visualization mappings as necessary. The session concluded with open discussions on additional system operations and compound interactions. All complete response sheets were collected at the end of the session.

\subsection{Data Analysis}
A total of $295$ complete interaction-visualization mappings were collected from the ideation workshop. After transcription, compound interactions were broken down and these mappings were parsed into their constituent interactions, yielding $314$ elemental statements. Two researchers then individually coded the elemental statements according to the initial set of interactions and commands. Throughout the process, an open-coding approach was employed for operations beyond the initial set, allowing the creation of new interaction codes as innovative strategies emerged during the workshop. The coding schemes were successively refined through iterative discussions between the two coders.

\section{Design Space of Tangible Cubes}

\begin{figure*}
    \centering
    \includegraphics[width=\textwidth]{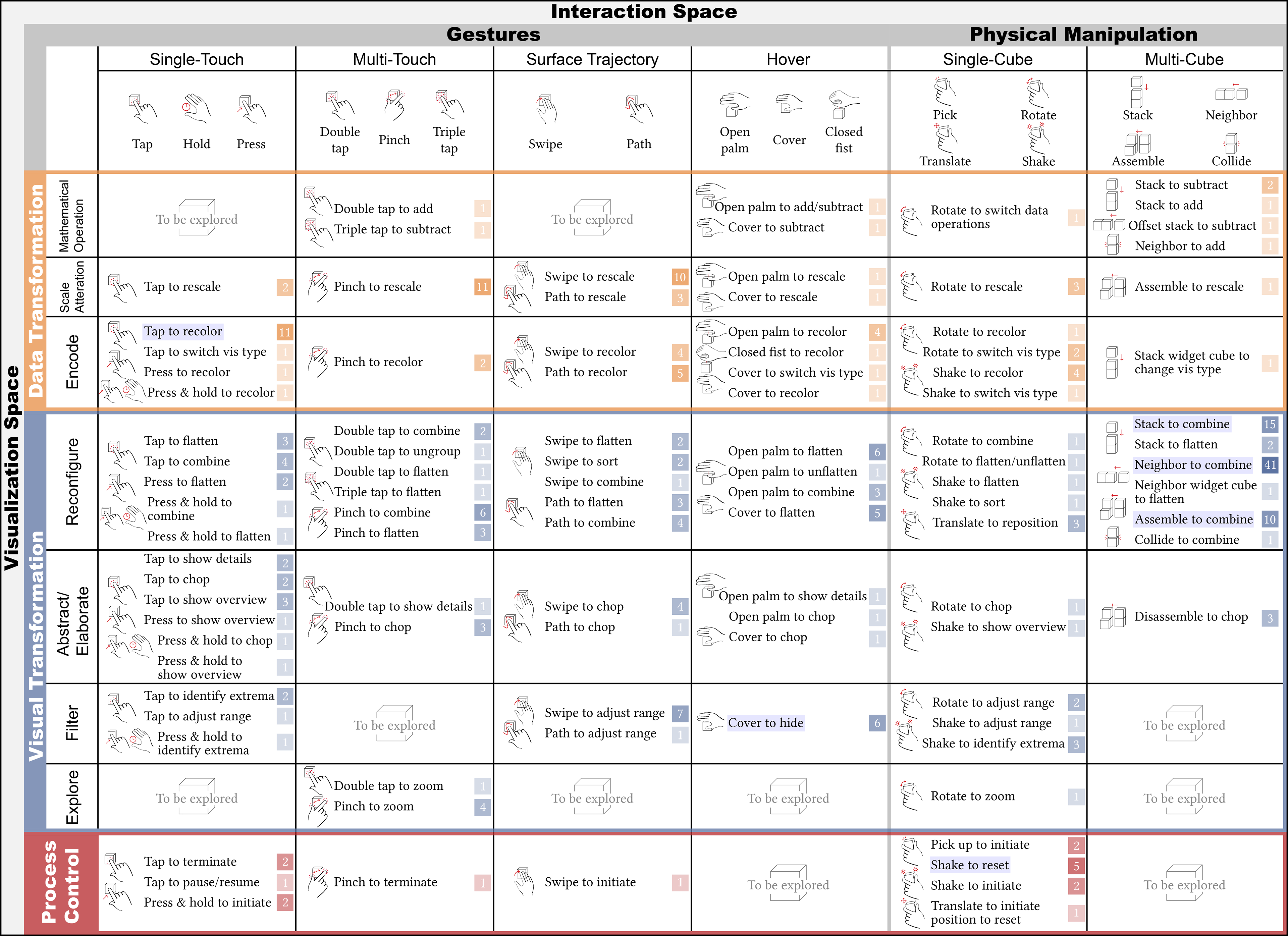}
    \caption{The design space of tangible cubes, structured along two dimensions: interaction space (encompassing gestures and physical manipulations) and visualization space (covering process control, visualization transformation, and data transformation). For each interaction-visualization pairing, multiple mapping combinations were suggested. The frequency of a specific mapping being proposed by the users is indicated by a color-coded square box \sqshade{ 7 }. The number within the box represents the frequency and the color intensity correlates to the frequency, with darker colors denoting higher frequency counts. 
     \highlightshade{Highlighting} denotes a mapping implemented in our prototype.}
    \Description{This design space image provides a comprehensive overview of interaction and visualization dimensions tailored for tangible cubes. Initially, the dimensions are divided into primary categories, each further refined into secondary categories for detailed interaction tasks. The interaction space is broadly categorized into two main modalities: 'Gestures' and 'Physical Manipulation'. Under 'Gestures', we see specific actions like 'Tap', 'Press', 'Double Tap', 'Pinch', 'Swipe', 'Hover', and their respective sub-actions, each indicating a specific outcome like 'recolor', 'rescale', 'flatten', etc. The 'Physical Manipulation' modality is further divided into 'Single-Cube' and 'Multi-Cube' interactions. 'Single-Cube' includes actions like 'Pick', 'Shake', 'Rotate', while 'Multi-Cube' encompasses 'Stack', 'Neighbor', 'Assemble', and 'Collide'. Each action and gesture is paired with a potential outcome or visualization change, such as 'switch data', 'combine', 'subtract', and more. The image serves as a detailed guide to understanding the myriad ways users can interact with tangible cubes and the resulting visual feedback in the MR environment.}
    \label{fig:designspace}
\end{figure*}

After analyzing the data from the ideation workshop, we constructed a design space for tangible cubes in spatio-temporal visualizations (\autoref{fig:designspace}), mapping user-identified interactions to their corresponding visual manipulations.

\vspace{2pt}
The interaction and visualization dimensions are first divided into primary categories tailored to specific tasks and targets. For finer granularity, we further subdivided each of the primary categories into secondary categories, providing a more detailed delineation of interaction tasks. The following sections offer in-depth explanations of these categories at each level.

\subsection{Interaction Space}
The interaction space defines the ways users engage with the tangible cubes. Broadly, this space can be categorized into two primary modalities: \textit{Gestures} and \textit{Physical Manipulation}. This classification illustrates the interaction techniques that are facilitated by the cube's physical properties and the user's instinctual responses.

\vspace{1em}
\begin{description}[nosep,leftmargin=1.5em,labelindent=0em,leftmargin=!,labelindent=!,itemindent=!,font=\normalfont\itshape]%
\item[\textbf{Gestures}]refer to interactions that are based on hand and finger movements. We further categorized gestures into four secondary categories: \textit{Single-Touch Gestures}, \textit{Multi-Touch Gestures}, \textit{Surface Trajectories} and \textit{Hover Gestures}.
\end{description}

\input{table/gestures}

\vspace{1em}
\begin{description}[nosep,leftmargin=1.5em,labelindent=0em,leftmargin=!,labelindent=!,itemindent=!,font=\normalfont\itshape]%
\item[\textbf{Physical Manipulations}]refer to the set of actions that involve moving, orienting and handling the tangible cubes themselves. These are further divided into secondary categories considering the multiplicity of cubes involved, including \textit{Single-Cube Manipulations} and \textit{Multi-Cube Manipulations}.
\end{description}

\input{table/physical_manipulations}

\subsection{Visualization Space}

The visualization space defines how visual information is presented and modified. It is divided into three primary categories based on the target of the commands: \textit{Data Transformation}, \textit{Visual Transformation}, and \textit{Process Control}. 

\begin{figure*}[htbp]
  \centering
  \includegraphics[width=\textwidth]{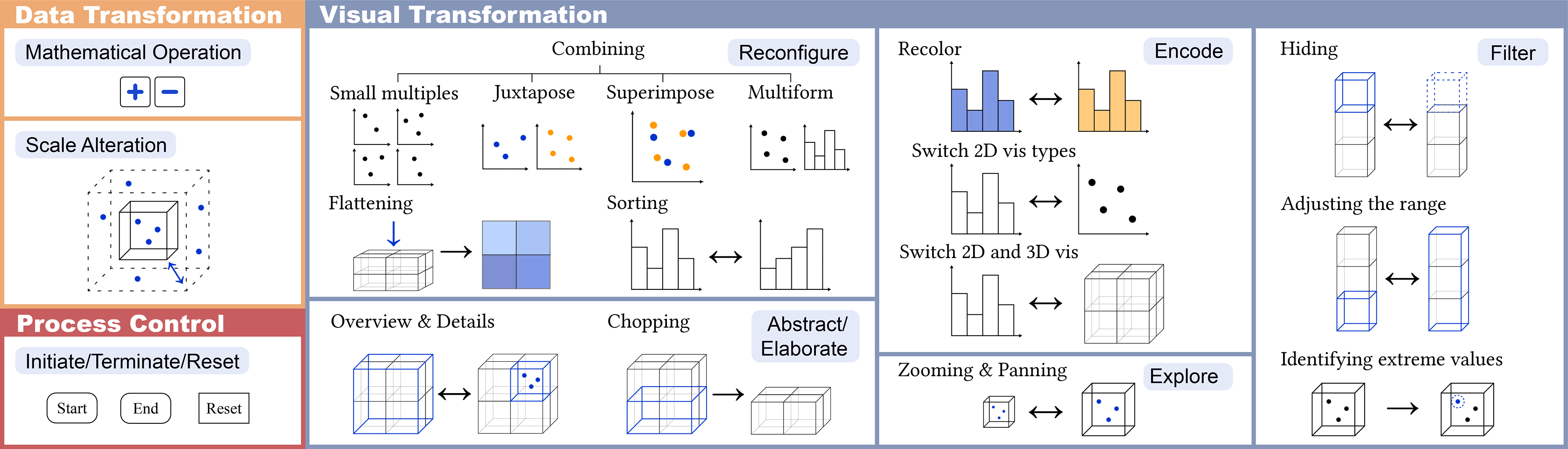}
  \caption{Pictorial summary of the commands in the visualization space, including three primary categories, namely data transformation, visual transformation and process control.}
  \Description{This visualization space image provides a structured representation of various data and visual transformations possible within a mixed reality environment. The image is organized into distinct categories and subcategories. Under 'Data Transformation', there are subcategories like 'Mathematical Operation' and 'Visual Transformation'. 'Visual Transformation' further breaks down into specific actions such as 'Combining', with techniques like 'Small multiples', 'Juxtapose', and 'Superimpose'. There's also 'Scale Alteration' which includes 'Reconfigure' and 'Multiform', and 'Flattening' which encompasses actions like 'Sorting'. 'Process Control' details actions like 'Initiate/Terminate/Reset', with specific commands like 'Start', 'End', and 'Reset'. The 'Overview & Details' section delves into more nuanced visual changes, including 'Chopping', 'Abstract/Elaborate', 'Recolor', and options to 'Switch 2D vis types' or between '2D and 3D vis'. Lastly, 'Zooming & Panning' covers actions to 'Encode', 'Explore', 'Hide', 'Filter', 'Adjust the range', and 'Identify extreme values'. The image serves as a comprehensive guide to understanding the range of visual transformations and interactions available in the MR visualization space.}
  \label{fig:Visualization_Space}
\end{figure*}

\vspace{1em}
\noindent
\DataTransformationShade{\textbf{Data Transformation}} pertains to changes in the underlying data represented by the tangible cubes. We identified two aspects of data transformation: \textit{Mathematical Operations} and \textit{Scale Alteration}.
\textit{Mathematical Operations} are arithmetic calculations, including \textit{addition} and \textit{subtraction}. \textit{Scale Alteration} refers to the process of \textit{rescaling} the range of data a cube embodies to support different levels of granularity or precision. 

\vspace{1em}
\noindent
\VisualTransformationShade{\textbf{Visual Transformation}} pertains to alterations in the visual depiction. We consolidated the visual commands identified by users, drawing upon the interaction taxonomies presented by Yi et al. \cite{Yi2007toward}, including \textit{Encode}, \textit{Reconfigure}, \textit{Abstract/Elaborate}, \textit{Filter} and \textit{Explore}.

\textbf{Encode} changes how the data is visually represented through operations such as \textit{recoloring} visual elements or \textit{switching visualization types} such as changing from a bar chart to a line chart.

\textbf{Reconfigure} involves the spatial rearrangement of data visualizations. This can be achieved through actions like \textit{combining}, \textit{ungrouping}, or \textit{sorting} visual elements to systematically organize data representations. Certain reconfigurations are uniquely tailored to 3D space-time cube constructs. For instance, \textit{flattening} a space-time cube along a designated axis can project its 3D visualization onto a 2D plane.

\textbf{Abstract/Elaborate} adjusts the level of abstraction and details. Unlike scale alteration, which expands or crops the actual data, this process does not change the underlying data. Instead, it shifts our perspective or the granularity of the details we view, allowing for a high-level \textit{overview} or in-depth examination of \textit{details} as needed. Specific to space-time cubes, a detailed view can be obtained through commands such as \textit{chopping}, which refers to segmenting the visualization into smaller components.

\textbf{Filter} selectively presents criteria-specific visual elements. For instance, \textit{adjusting the range} of data along a specific axis, \textit{identifying extreme values} or \textit{hiding} certain visual elements.

\textbf{Explore} overcomes display space constraints, enabling users to uncover new insights via \textit{zooming and panning}.

\vspace{0.6em}
\noindent
\ProcessControlShade{\textbf{Process Control}}
Process control commands are system specifications on the life cycle of the operations. For instance, to \textit{initiate} or \textit{terminate} an operation or the program, or to \textit{reset} the visualization to its original state.

\subsection{Utilizing the Design Space}
\label{utilizing}
While the design space showcases a wide array of possible mappings, it entails redundancies where one interaction action could potentially map to multiple visualization commands. However, in practical embodied interaction, a single interaction is typically devoted to one specific visualization command to avoid ambiguity. For instance, if \textit{pinch} is used to \textit{rescale} the visualization, it cannot be used for \textit{recoloring} anymore. This implies the necessity for designers to prioritize: the most intuitive, expressive and effective interactions should be reserved for the most critical visualization tasks depending on the specific context. To illustrate the practical utility of the design space, we provide an example set of mutually exclusive, one-to-one interaction-visualization pairings drawn from the design space (\autoref{tab:utility}).

\input{table/use_design_space}

\section{Prototyping From The Design Space}
The design space provides a promising foundation for the tangible exploration of spatio-temporal data. However, the practical realization and implementation of these operations bear a crucial impact on both the user experience and the effectiveness of the visualizations. Therefore, we demonstrate the practicality of implementing these interactions and further validate the identified pairings in the design space by creating a proof-of-concept prototype. We selected and implemented a small subset of interaction mappings (highlighted in \autoref{tab:utility}) for the spatio-temporal use case (section \ref{subsec: datacontext}) described in the ideation workshop.

\begin{figure}[ht]
  \centering
  \includegraphics[width=0.8\columnwidth]{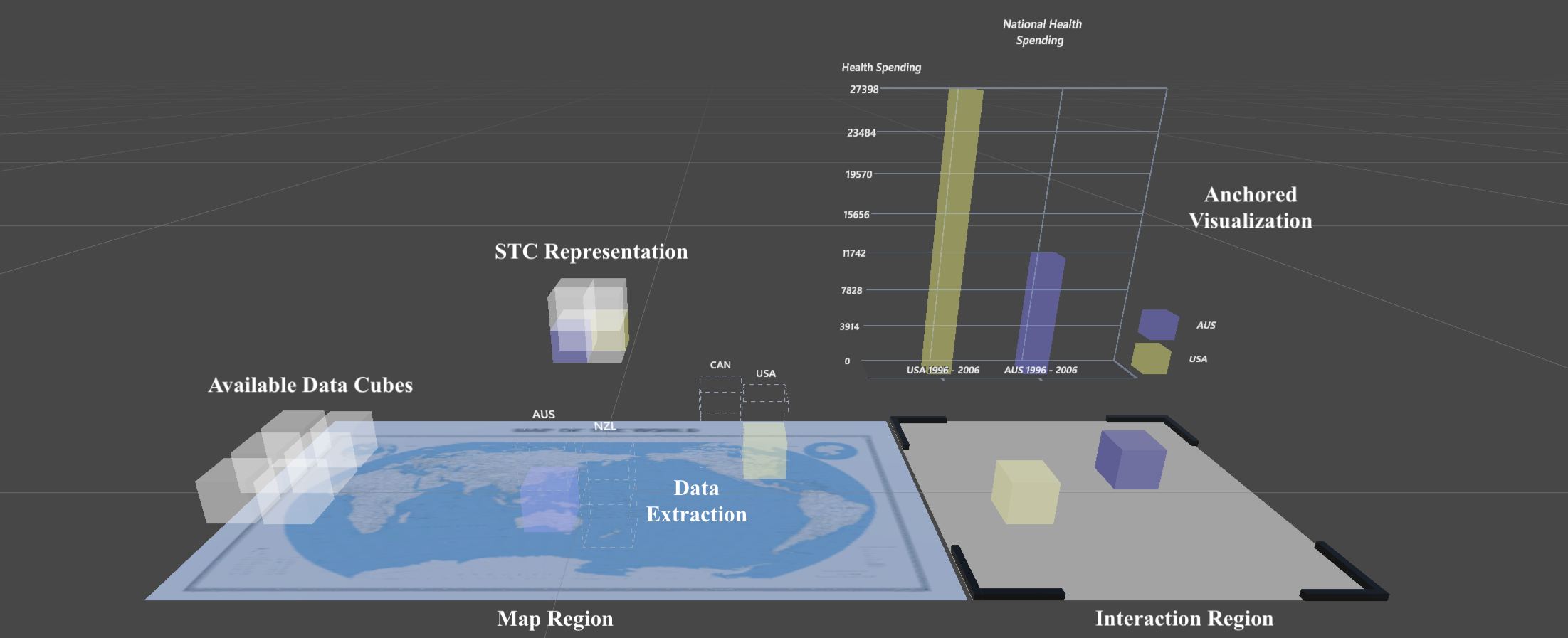}
  \caption{Illustration of the prototype interface. The interface consists of four components: data cubes, a map region, an interaction region, and visualizations.}
  \Description{An illustration showcasing the prototype interface developed in Unity 3D. The interface is composed of four distinct components. First, there are 'data cubes' which appear as three-dimensional blocks representing different sets of data. Adjacent to the data cubes is the 'map region', a designated area that provides a geographical context for the data. The 'interaction region' is a specific zone where users can engage with and manipulate the data cubes to trigger various visualizations. Lastly, the 'visualizations' are dynamic graphical representations that change based on user interactions with the data cubes, providing insights and patterns from the data. The entire interface is designed to offer an immersive and interactive experience for users to explore and understand the data in a mixed reality environment.}
  \label{fig:MR_interface}
\end{figure}

\subsection{The MR Interface}
Our interface design is categorized into four functional components: tangible cubes serving as embodied data carriers, a map region providing spatial context, an interaction area for manipulating the tangible cubes, and visualizations that adapt based on user interactions. \autoref{fig:MR_interface} provides an overview of the interface.

\vspace{1em}

\noindent
\includegraphics[height=1em]{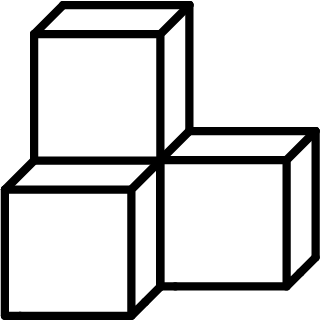} \textbf{Tangible Cubes: Physical Data Carriers}
 
\begin{wrapfigure}{l}{0.35\textwidth} 
  \centering
  \includegraphics[width=0.3\columnwidth]{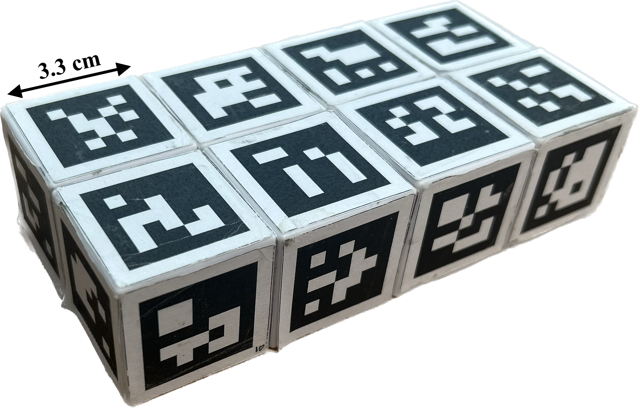}
  \caption{Tangible cubes used.}
  \par
  \label{fig:tangible_cubes}
\end{wrapfigure}

We employed a set of eight plastic cubes, each with an edge length of $3.3$cm and weight of approximately $15$g. These compact cubes can be manipulated either as standalone units or collectively as a cohesive group. For a more comprehensive representation in the context of a space-time cube ensemble, these individual cubes can be seamlessly assembled into a larger $2\times2\times2$ configuration, resulting in a medium-sized cube with an edge length of $6.6$cm. The size of the cube was selected to facilitate easy manipulation and combination. Additionally, the cubes were magnetic to ensure stability and facilitate the creation of ensembles.

\vspace{1em}
\noindent
\includegraphics[height=1em]{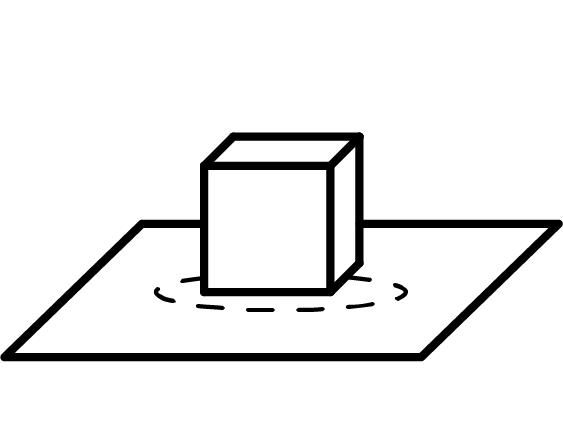} \textbf{Map Region: Spatial Context}

\begin{wrapfigure}{l}{0.35\textwidth}
  \centering
  \includegraphics[width=0.32\columnwidth]{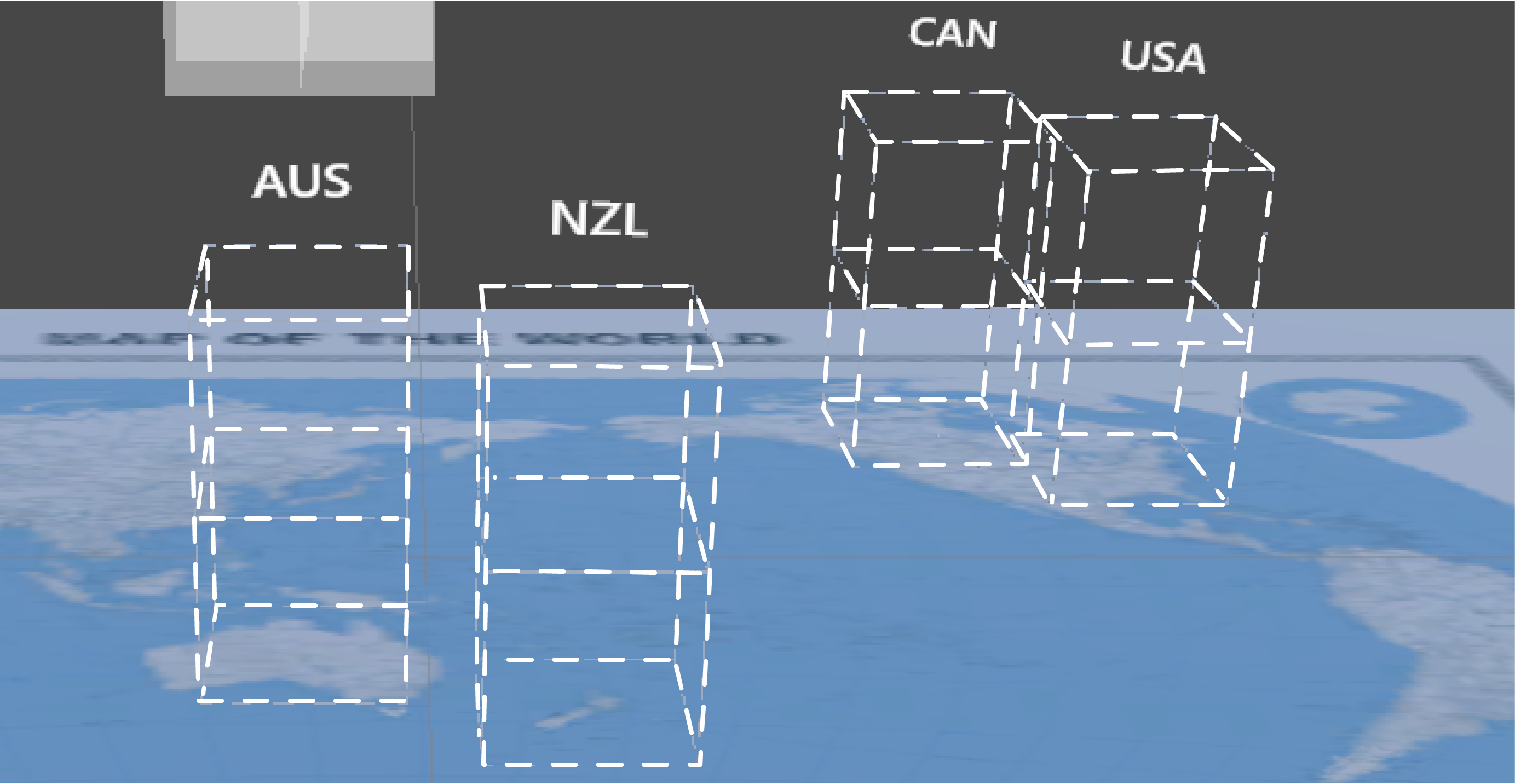}
  \includegraphics[width=0.32\columnwidth]{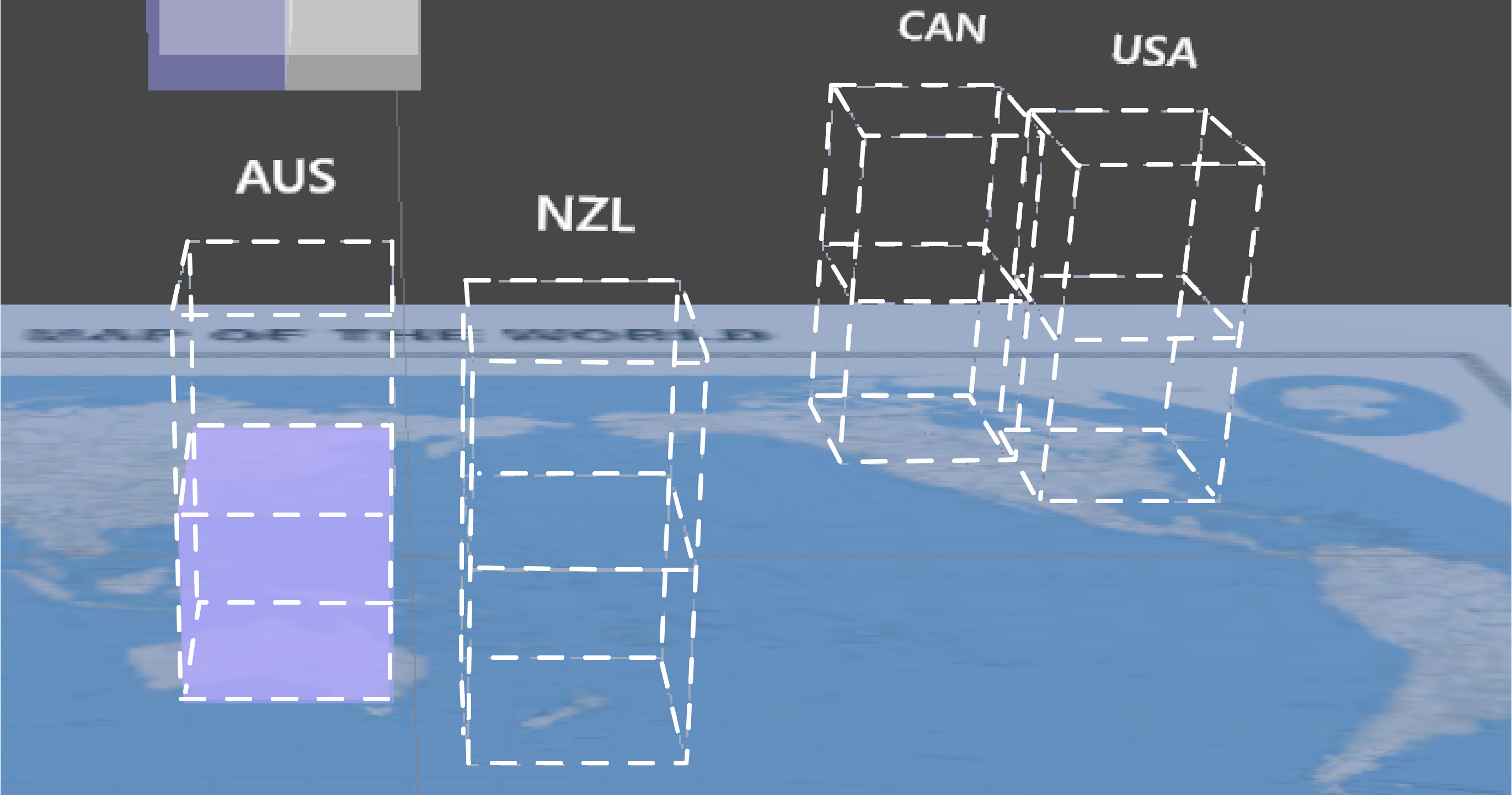}
  \caption{Illustration of the map region.}
  \par
  \label{fig:MR_map}
\end{wrapfigure}

The map region within our prototype establishes a spatial context for the dataset. As users engage with the system, they encounter virtual cubes with dashed outlines that symbolize national health spending data (\autoref{fig:MR_map} Top). These cubes are strategically positioned on the map, mirroring their real-world geographical locations. This design choice allows users to correlate specific cubes with their respective countries. The map region also serves as the foundation for data extraction: users align the tangible cubes with their virtual counterparts in the MR environment. Once aligned, a connection is established between the two and the data represented within the virtual cubes are embodied in the tangibles. To provide users with a visual confirmation of this transfer, each virtual cube illuminates in a distinct color (\autoref{fig:MR_map} Bottom), representative of its associated country. Consequently, the tangible cubes evolve into data-bearing entities ready for subsequent interactions.

\vspace{1em}
\noindent
\includegraphics[height=1em]{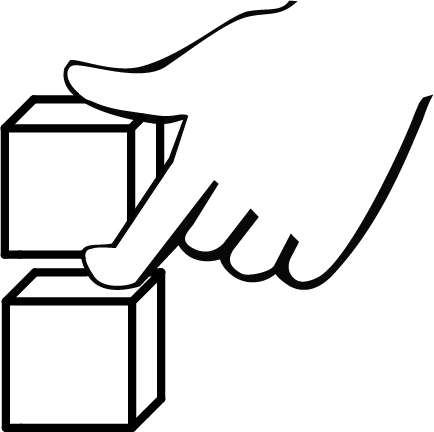} \textbf{Interaction Region:
Tangible Interactions}

\begin{wrapfigure}{l}{0.35\textwidth}
  \centering
  \includegraphics[width=0.32\columnwidth]{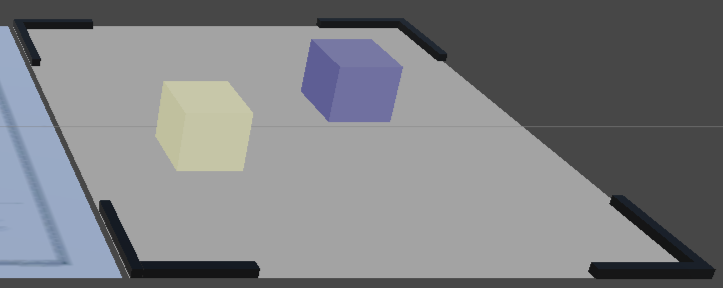}
  \caption{The interaction region.}
  \par
  \label{fig:MR_interaction}
\end{wrapfigure}

The interaction region serves as a dedicated zone where users can actively engage with the tangible cubes. Visually represented as a subtle gray square, this region is strategically situated adjacent to the right of the map region. Such positioning ensures a swift transition between these two core functional areas. Once data is extracted, users can manipulate the system by transferring cubes from the map region to the interaction region, where a set of desired interactions can be executed to further explore the data.

\vspace{1.5em}
\noindent
\includegraphics[height=1em]{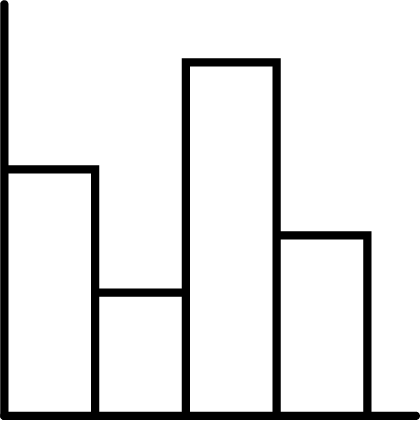} \textbf{Visualizations: Chart Structure and Dynamics}
\vspace{0.5em}

User interactions directly influence the presentation and modification of data visualizations. For our proof-of-concept prototype, we employed bar charts of distinct \textit{structures} and \textit{dynamism of display} to reflect the changes in the combination of cubes while offering different means for observation.

\begin{figure}[htbp]
  \centering
  \includegraphics[width=\columnwidth]{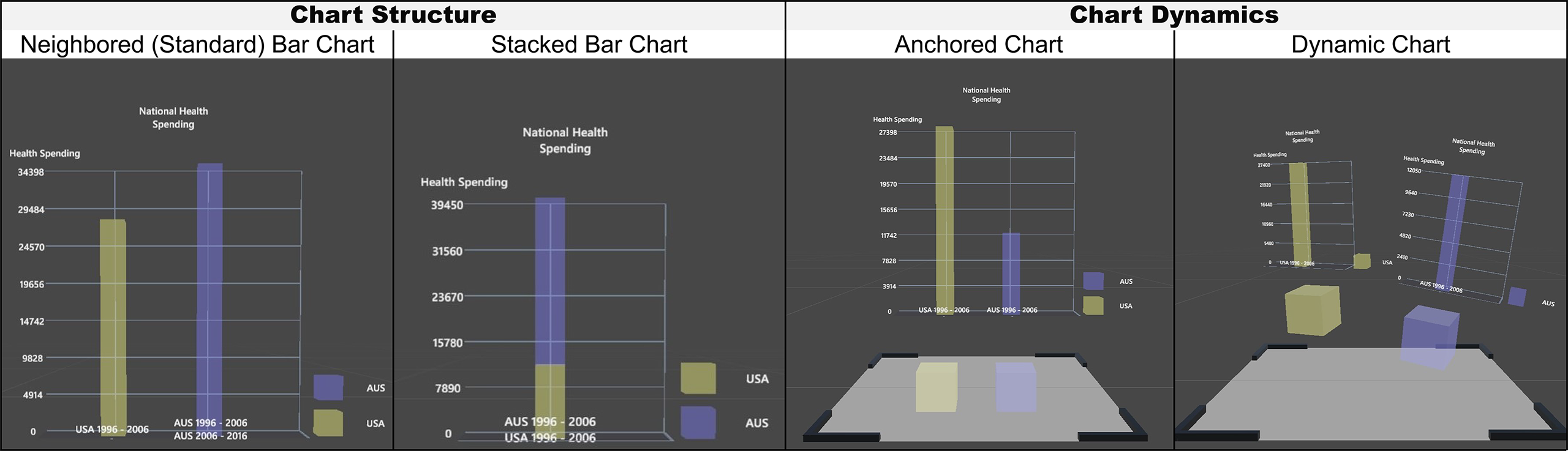}
  \caption{Depiction of bar chart variations, differentiated by structure and display dynamics. A: Neighbored charts displayed by default and in response to neighboring cube actions; B: Stacked bar charts displayed when cubes are stacked. C: Anchored charts fixed behind the interaction region for consistent referencing; D: Dynamic charts displayed on top of cube surface, moving in tandem with them.}
  \Description{A visual representation of various bar chart variations against a black background. The image is divided into four sections, each illustrating a different chart style. In section 'A', yellow and purple bars are displayed side by side, representing 'Neighbored charts' that appear by default and in response to neighboring cube actions. Section 'B' showcases 'Stacked bar charts', where the yellow and purple bars are vertically stacked, symbolizing the action of stacking cubes. In section 'C', bars are shown behind a designated 'interaction region', representing 'Anchored charts' that maintain a consistent position for easy referencing. Lastly, section 'D' presents 'Dynamic charts', where the bars are displayed directly on the top surface of corresponding yellow and purple cubes, moving in sync with the cubes' movements. Each variation offers a unique way of visualizing data in relation to tangible cube interactions.}
  \label{fig:MR_barchart}
\end{figure}

\vspace{2pt}
There were two types of chart structures used in the prototype: neighbored (standard) bar charts (\autoref{fig:MR_barchart}A) and stacked bar charts (\autoref{fig:MR_barchart}B). These are designed to reflect the combination of the tangible cubes. For example, when two tangible cubes are \textit{neighbored}, the neighbored bar chart visuals will be displayed. Conversely, when two tangible cubes are \textit{stacked}, the stacked bar chart visuals will be displayed.

\vspace{2pt}
On the other hand, to explore the adaptability of visualizations and their impact on user experience, we introduced two distinct dynamics of visualization styles: \textit{Anchored visualizations} (\autoref{fig:MR_barchart}C) and \textit{Dynamic visualizations} (\autoref{fig:MR_barchart}D). Anchored visuals remain fixed behind the interaction area, providing a stable point of reference. These anchored charts refresh each time the cubes return to the interaction surface. Conversely, the dynamic visuals appear atop the tangible cubes, adjusting in real time to the cube's movement and orientation. They become visible when users lift the cubes from the interaction surface, offering an immersive, hands-on experience. In essence, our system combines the immediacy of dynamic visuals with the clarity of anchored visuals, catering to both interactive exploration and consistent referencing.

\vspace{1em}

\subsection{Interaction-Visualization Pairs}

\label{sec:interaction_pair}
\begin{figure*}[ht]
  \centering
  \includegraphics[width=\linewidth]{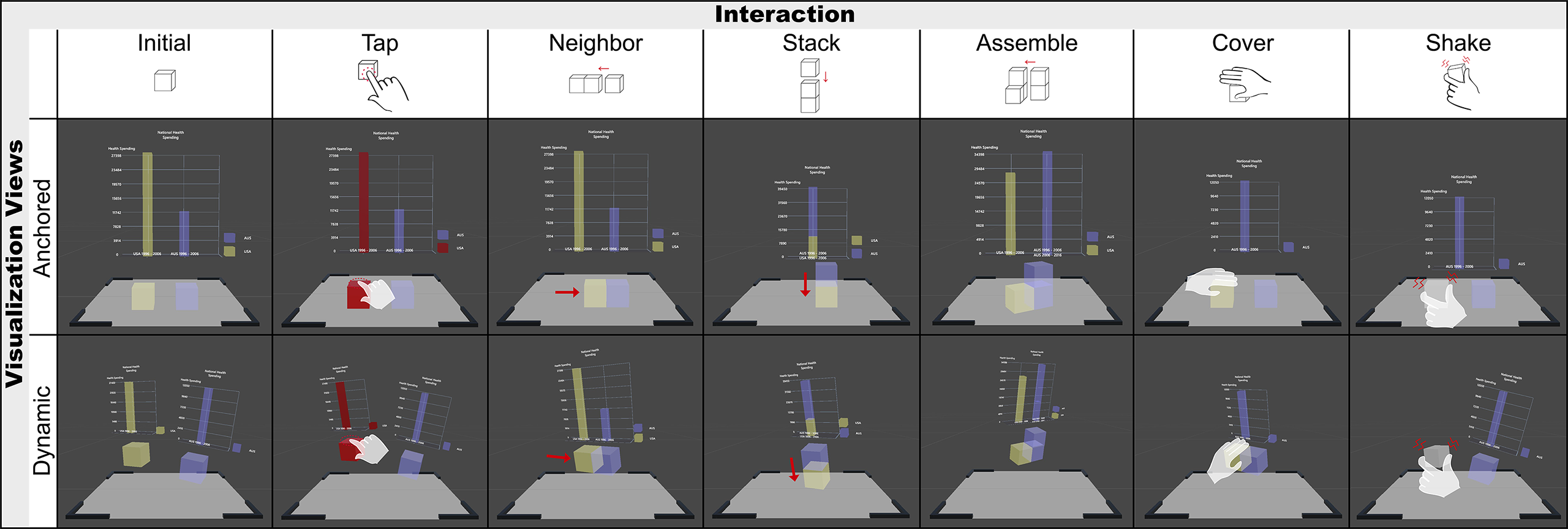}
    \caption{Illustration of visualization adaptations based on interactions with two tangible cubes within the interaction region. The top row showcases the anchored visualization perspective, while the bottom row presents the dynamic visualization perspective. The first column displays the initial visualization state prior to any interactions. Columns two through seven depict the visualization responses to specific tangible cube interactions. For demonstration clarity, all actions were executed on the yellow (left) cube.}
  \label{fig:MR_interactionvisualization}
\end{figure*}

In the interaction region, we implemented four visualization tasks: \textit{Encode}, \textit{Reconfigure}, \textit{Filter}, and \textit{Process Control}, as detailed in the design space (shaded in \autoref{tab:utility}). These tasks are illustrated in \autoref{fig:MR_interactionvisualization}.

\vspace{5pt}
\begin{description}[nosep,leftmargin=1.5em,labelindent=0em,leftmargin=!,labelindent=!,itemindent=!,font=\normalfont\itshape]%
\item[\textbf{\textit{Encode: Tap \includegraphics[height=1em]{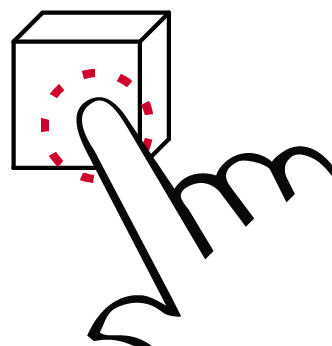}} $\rightarrow$ Recolor \includegraphics[height=1em]{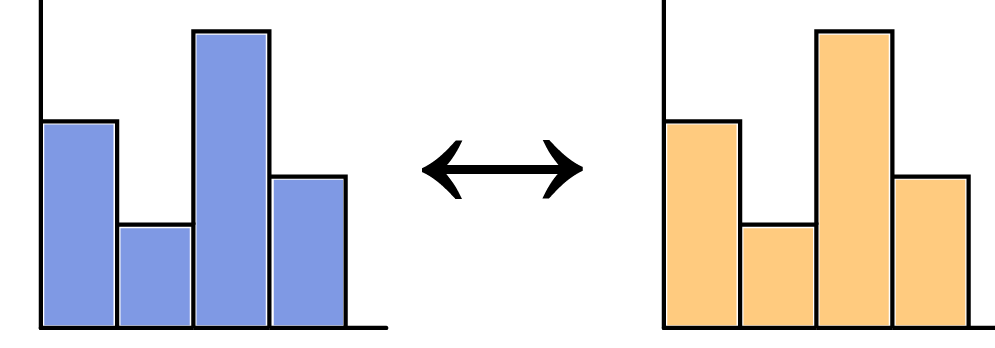}}.] A \textit{tap} on the surface of a tangible cube triggers a \textit{recolor} operation in its associated visualizations, changing the color of the corresponding bar. This can apply to a single cube or a specific cube within an assembled structure, allowing the users to concentrate on the country or time span of their interest.
\end{description}

\vspace{2pt}
\begin{description}[nosep,leftmargin=1.5em,labelindent=0em,leftmargin=!,labelindent=!,itemindent=!,font=\normalfont\itshape]%
\item[\textbf{\textit{Reconfig: Neighbor, Stack, Assemble \includegraphics[height=1em]{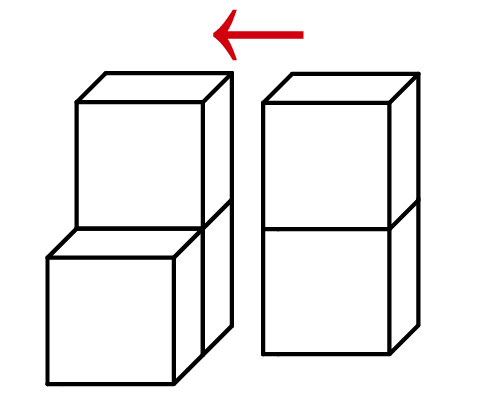} $\rightarrow$ Combine \includegraphics[height=1em]{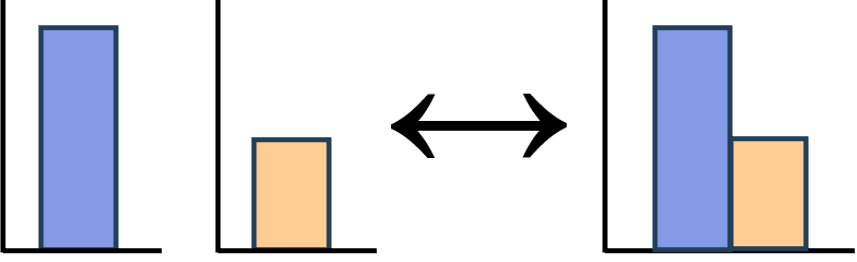}.}}] 
Users can combine tangible cubes through three distinct operations: \textit{neighboring}, \textit{stacking}, and \textit{assembling}. When individual cubes are \textit{neighbored}, their dynamic bar charts converge and merge into a unified bar chart, displayed at the the midpoint of the two cubes' top surfaces. This unified chart is then reflected in the anchored chart when the cubes are returned to the interaction surface. When individual cubes are \textit{stacked}, the visualizations transition from standard bar charts to stacked bar charts. \textit{Assemble} acts as the combination of neighboring and stacking, hence transforming the bar charts accordingly.
\end{description}

\begin{description}[nosep,leftmargin=1.5em,labelindent=0em,leftmargin=!,labelindent=!,itemindent=!,font=\normalfont\itshape]%
\item[\textbf{\textit{Filter: Cover \includegraphics[height=1em]{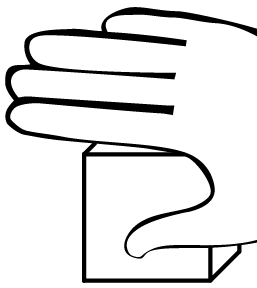} $\rightarrow$ Hide \includegraphics[height=1em]{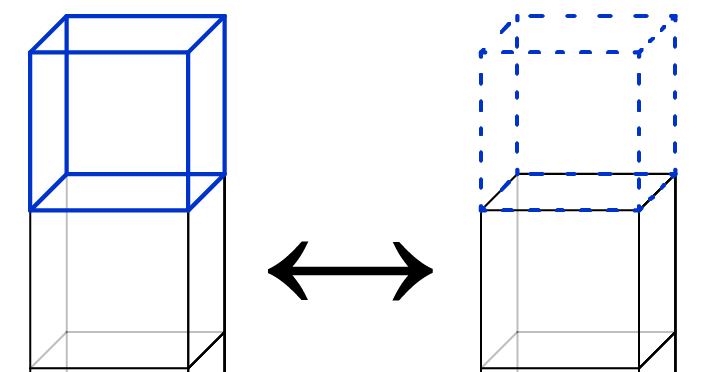}.}}] Once the tangible cubes are combined, the visibility of the data can be controlled. By \textit{covering} one or a subset of cubes, users can hide the underlying data from the visualization. For instance, the bars corresponding to the covered cubes will disappear in the rendered visualizations. Once the tangible cubes are uncovered, the data re-appear on the chart. This operation manages the complexity of the visualization and reserves the most relevant data elements for the users' investigation.
\end{description}

\begin{description}[nosep,leftmargin=1.5em,labelindent=0em,leftmargin=!,labelindent=!,itemindent=!,font=\normalfont\itshape]%
\item[\textbf{\textit{Process Control: Shake \includegraphics[height=1em]{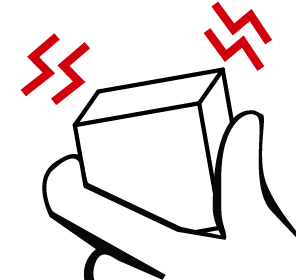} $\rightarrow$ Reset \includegraphics[height=1em]{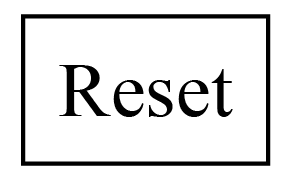}}}.] To accommodate the iterative nature of data exploration, we have integrated a process control mechanism. After users complete their current visualization tasks, a simple \textit{shake} of any tangible cube clears its associated visualizations in the charts and detaches the associated data. Once reset, the tangible cube can then be used to extract new data. This enables the transition between different data exploration sessions and prepares the system for the next round of data exploration.
\end{description}

\subsection{Implementation}
The MR environment, which encompasses virtual cubes, a map region and an interaction region, is rendered using Unity. To create the visualizations, we transformed the pipeline of a Unity 2D visualization tool, E2Chart \cite{e2chart2023}, to generate the 3D bar chart representations within this environment. To ensure a robust and controlled testing environment, we employed the Wizard of Oz (WOz) evaluation technique. This method was chosen because it allows for more flexible and adaptive responses to user behaviors. 
\begin{edited}
From the participant's perspective, they freely picked up the initialized tangible cubes as data carriers for associating any data set upon engaging in the map region. From the investigator's perspective, they closely monitored the user's interactions with the cube, identifying specific actions, and then manually initiated the corresponding visualization commands using pre-configured inputs on a controller. 
\end{edited}
 To allow both the investigator and the participant to observe the MR environment simultaneously, we synchronized two Microsoft Hololens2 head-mounted displays using WebSocket \cite{websocketapi2021}. Prior to each session, the MR interface's position was calibrated to ensure optimal visual synchrony.

\section{Evaluation}
\begin{figure*}
  \centering
\includegraphics[width=\linewidth]{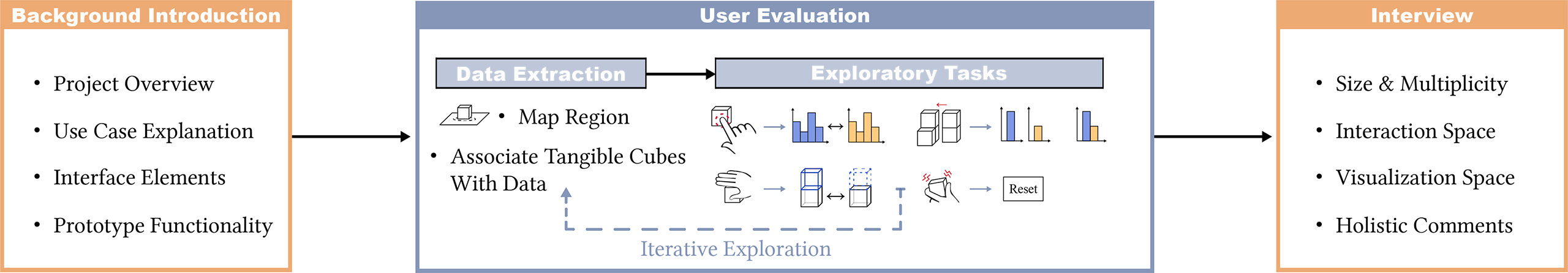}
  \caption{Flowchart of the user evaluation process.}
  \Description{A flowchart detailing the user evaluation process, set against a white background. The flowchart is organized sequentially, starting with 'Background Introduction' at the left. This section includes sub-points like 'Project Overview', 'Use Case Explanation', 'Interface Elements', and 'Prototype Functionality'. Following this, the 'User Evaluation' phase is introduced. The first step in user evaluation is 'Data Extraction', which covers 'Map region' and 'Associate tangible Cubes with Data'. leading to specific tasks such as 'Exploratory Tasks' and 'Reset'. The process then moves to 'Iterative Exploration' and concludes with an 'Interview' section. The interview covers feedback areas like 'Size & Multiplicity', 'Interaction Space', 'Visualization Space', and 'Holistic Comments'. The flowchart provides a structured overview of the entire user evaluation journey.}
  \label{fig:userstudy_flow}
\end{figure*}

To assess the practicality of our proposed interactions and the functionality of the prototype, we designed a qualitative user evaluation approved by our university ethics committee. \autoref{fig:userstudy_flow} depicts the user evaluation process.

\subsection{Participants}
We invited six participants to take part in the evaluation. To obtain a holistic perspective, the participants came from two cohorts: three of the participants (P1-P3) were HCI/MR experts (two lecturers, one senior PhD student), and the other three participants (P4-P6) were intermediate users with some knowledge but limited hands-on experience in MR and tangibles (one undergraduate student, two master's students). This combination aimed to gather both professional feedback regarding data exploration and insights into the intuitiveness and engagement of our prototype system.

\subsection{User Evaluation Procedure}
Participants were first introduced to the project background and the use case (section \ref{subsec: datacontext}), then given an overview of the prototype's purpose and functionality. Subsequently, they were provided with the opportunity to familiarize themselves with the tangible cubes and the associated interactions (\autoref{tab:utility}). As practice, they performed the actions outlined in the task using the tangible cubes and were encouraged to raise questions as needed. Meanwhile, a facilitator closely monitored the practice session to offer guidance.

Once participants fully understood the interface, the interactions and their corresponding visualization mappings, they were guided to engage with the prototype. Specifically, they were instructed to extract data from the map region and to interact with the tangible cubes in the interaction region. They performed the tasks outlined in section \ref{sec:interaction_pair}, including \textit{Encode}, \textit{Reconfigure}, \textit{Filter} and \textit{Process Control}. These tasks were executed in varying sequences to align with each participant's unique exploratory interests. Throughout the process, participants were encouraged to think aloud and verbalize their thoughts and observations, providing us with insights into their intentions and experiences.

After all tasks were completed, an interview was conducted to gauge the participants' experience. The interview consisted of a mix of open-ended and specific questions (see \autoref{app:appendix2}), designed to provide a well-rounded understanding of the size and multiplicity of the cubes as well as the effectiveness of the tangible interaction and visualization.

\subsection{Results}
Overall, participants were pleased with their experience, suggesting the viability of the tangible cube-based design space for data visualization. Their feedback not only validated the conceptual foundation of the design but also provided insights into the prototype's specific strengths and areas for enhancement.

\subsubsection{Optimal Size and Multiplicity}
Most participants found the size of the tangible cubes to be appropriate for the tasks, supporting ease of manipulation and maneuverability. For instance, P1 noted that ``the size of the cube is just right,'' and P3 echoed ``the cubes are easy to handle and manipulate.'' However, while there is a general consensus on the objective ease of manipulation of the tangible cubes, concerns were raised about the relative size of the cubes in relation to the interface environment, particularly the geographical regions within the map region. P4 expressed, ``Sometimes the cube feels a little too big for certain regions on the map, and it kind of occludes the underlying countries, which makes them hard to see.'' Similarly, P5 mentioned that the size of the cubes was suitable for the current tasks, but also hinted at potential challenges in more complex scenarios, ``I imagine if there were more cubes, it might be a bit troublesome in a smaller scene.'' In addition to ergonomic and interaction considerations, these feedbacks underscore the importance of balancing the tangible cube size with the complexity and scale of the data visualization tasks.

Regarding multiplicity, all participants agreed that the availability of multiple cubes enhanced the data exploration experience. For instance, P5's comment that the quantity of the cubes is ``just right'' was a sentiment echoed by others. One of the primary advantages of using multiple cubes is its ability to perform comparative analysis. As P2 noted, ``having multiple cubes is useful, especially when comparing data from different countries.'' Having two or more cubes enabled the users to physically juxtapose data points, making comparisons more intuitive. Moreover, multiple cubes support iterative data exploration. During the evaluation, users often wanted to go back and forth between different data points and compare new data with previously explored data. Instead of resetting or reconfiguring a single cube to view different data points, our system allowed users to take a new cube for new data. P6 suggested that ``using a new cube might be better'' when there is a lot of data to explore. In this case, having multiple cubes allowed users to keep some data points constant while changing others, facilitating iterative exploration.

\subsubsection{Physicality Enhances Intuitiveness}
The tangible nature of the cubes was applauded by the participants for its intuitiveness and immediacy. As P2 and P3 mentioned, the tangible interactions are rooted in ``everyday experiences''. Translating these familiar actions into data exploration makes the process more intuitive. Using tangibles was also thought to enhance the directness of interaction. As P4 noted, the tangible interaction ``feels more direct''. Likewise, P6 mentioned that ``holding something directly in hand feels good.'' When users can touch, move and manipulate data representations, they perceive and observe the data in a more concrete way. 

One interesting comment is that the tangible nature of the cubes instilled a sense of confidence in the users. P4 hinted at this increased confidence by suggesting that the tangibility of the cubes made the data exploration feel more grounded and reliable. They emphasized that the haptic feedback from ``grabbing'' the tangible cubes complemented the visual feedback of the cube's location, providing more trust in where the cube actually lies within the MR environment.

\subsubsection{Balancing Dynamic Exploration with Anchored Clarity}
The dynamic and anchored visualizations were designed to offer users a balance between exploratory freedom and clarity.

During the evaluation, the dynamic visualizations, which moved with the tangible cubes, were described as engaging and novel. For instance, P2 commented that ``the dynamic charts are fun.'' This sentiment of novelty was also captured by P5 who found this visualization style to be embodied and an interesting departure from traditional static visualizations. However, the very nature of their dynamism led to challenges in consistent data interpretation. For instance, the constant movement and reorientation could occasionally make it difficult for users to track specific data points. In addition, P1, P4 and P6 all mentioned the notion that they were unable to see the details in the dynamic charts due to the limited size of the rendering, and mainly relied on the anchored chart for data understanding.

The anchored visualizations were favored due to their predictability, clarity, and stability. They consistently appeared in a fixed position, featured aligned charts that were larger in size, and remained stationary. These characteristics made them a reliable reference, enabling participants to confidently compare bars and extract insights. P4's observation, ``the anchored charts give me a stable reference,'' underscores the significance of having a well-aligned, stationary visualization, particularly when navigating multiple data points.

While some participants (P1, P4, P6) had a clear preference for the anchored visualizations over the dynamic ones, the rest appreciated the combination of both. This dual-style approach allowed users to engage in freeform exploration with the dynamic views while still having anchored charts as a reliable reference for more in-depth insights. P3's feedback encapsulates this balance: ``appreciated the dynamic charts for exploration but relied on the anchored charts for understanding.'' In essence, the prototype's dual visualization approach offers an engaging way to interact with data through dynamic charts, but it is the anchored visualizations that ensure users have a clear, stable reference to ground their insights and interpretations.

In summary, our evaluation results underscore the feasibility and potential of the tangible cube-based design space for data visualization. Participants, both experts and intermediates, appreciated the tangibility of the cubes, emphasizing their intuitive nature and the directness of interaction they offer. The interaction-to-visualization mappings we identified resonated with users' expectations, making the data exploration process more intuitive and grounded. A highlight of the study was the combination of cubes, which facilitated comparative analysis and iterative exploration, enhancing the overall data exploration experience.

In terms of visualization styles, our focus on both dynamic and anchored visualizations revealed the importance of striking a balance. While the dynamic visualizations brought novelty and embodied interaction to the forefront, the anchored visualizations provided clarity, stability, and predictability, serving as a reliable reference point. This balance between dynamism and clarity ensures that users can engage in exploratory data interactions while still having a clear frame of reference to derive meaningful insights.

\section{Discussion}

The prototype served as a tangible manifestation of the design space we envisioned. Through its implementation and the subsequent user evaluation, we were able to glean insights that not only validate the feasibility of the design space but also highlight areas of refinement and future exploration. Here, we weave the insights into a broader discussion of the significance and potential utility of the design space for future designers.

\begin{figure*}[ht]
  \centering
  \includegraphics[width=\linewidth]{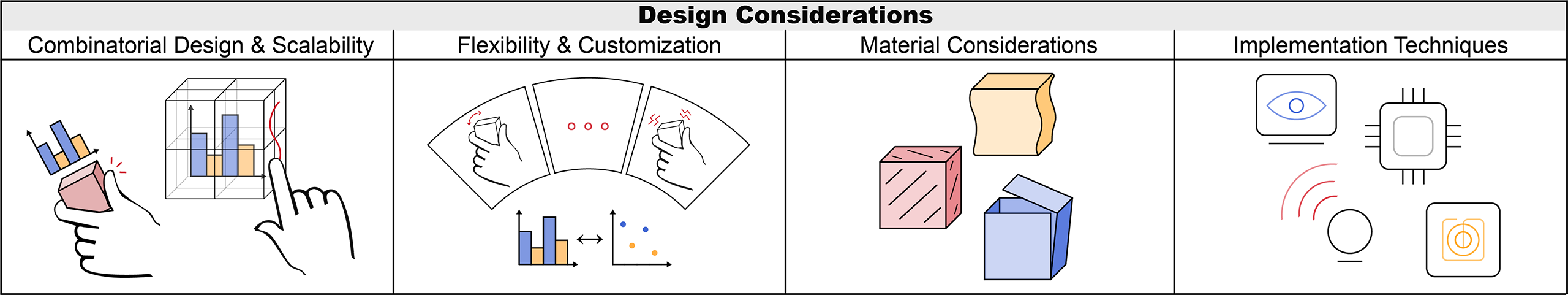}
  \caption{A visual representation of the multifaceted design considerations: (1) potential of combinatorial design and scalability, (2) flexibility and customization, (3) diverse tactile experience and interaction dynamics offered by different materials, and (4) icons representing various implementation techniques.}
    \Description{A visual diagram illustrating four key design considerations for tangible MR systems. The image is divided into four sections, each representing a distinct aspect of design: 1. Combinatorial Design and Scalability: Depicts an array of interconnected cubes, symbolizing the vast potential of combining different sizes and multiplicities for diverse interactions and visualizations. 2. Flexibility and Customization: Features a cube with adjustable sliders and dials, emphasizing the adaptability of the system to user preferences and the importance of personalization. 3. Material Considerations: Shows cubes made of different textures and patterns, highlighting the tactile experiences and interaction dynamics that various materials can offer. 4. Implementation Techniques: Displays a series of technical icons, representing the tools and methods used in tracking and interaction within tangible MR environments. Each section is visually distinct, and the overall image conveys the multifaceted nature of design considerations in the realm of tangible MR systems.}
  \label{fig:design_considerations}
\end{figure*}

\subsection{Combinatorial Design and Scalability}
Our design space outlines a range of interactions that users can perform with multiple small tangible cubes. These cubes can also be combined to form larger, cohesive structures, offering unique and promising avenues for more versatile visualization strategies and scaling up data exploration.

As we discussed in \autoref{sec:taxonomy}, the sizes and multiplicity of the cubes present both opportunities and constraints in designing the location and style of interaction and visualization. When smaller cubes are combined into larger structures, their visual strategies and interaction techniques can also evolve with the expanding size. Different visual and interactive techniques can be designed for manipulating individual cubes versus larger combined structures. 

For example, in the interaction space, designers could adopt single-touch gestures for interacting with individual cubes, given their small volume. These simple gestures are well-suited for the limited surface area of a small cube. In contrast, when cubes are combined into a larger structure, the ample surface area allows for the design of more complex surface trajectories. As suggested by P4 and P5 during our user study, gestures could also be used on distinct faces of the cubes to trigger different visualization commands. 

Similarly for the visualization space, while our prototype offered two visualization styles (dynamic and anchored) that are consistently used for both individual cube manipulations and combinatorial structures, a wide range of different visualization styles and locations are available for consideration. According to our categorization, dynamic charts are displayed ``\textit{above}'' the tangible cubes, whereas anchored charts are displayed to the ``\textit{side}''. These are common spatial locations for small cubes (\autoref{tab:summary}). However, when scaling up and combining cubes into larger structures, additional visualization spaces become available. For instance, once a combined structure reaches a certain size, the internal volume could be used for ``\textit{inside}'' visualizations, offering additional adaptive design considerations. 

However, scaling up the number of cubes introduces its own set of challenges. Participants in our user study expressed concerns that having ``too many'' cubes could increase cognitive load and potentially cause confusion. To mitigate this, designers should consider implementing features that facilitate user interaction while minimizing memorability demands. Effective color-coding systems that easily segregate data points into different clusters and labelling strategies that remind users of data provenance could be useful. Ultimately, a balance must be struck between the complexity of the data and the multiplicity of the tangible cubes. The cognitive burden and potential chaos caused by an overabundance of cubes should not overshadow the benefits of their tangibility.


\subsection{Flexibility and Customization}
Our design space is envisioned as a canvas of possibilities rather than a prescriptive set of mappings. It aims to capture the intuitive choices users might gravitate towards, offering a flexible framework that can be tailored to specific contexts and needs.

As highlighted in section \ref{utilizing}, while the design space provides a broad spectrum of possibilities, it is essential for designers to prioritize. The most intuitive mappings should align with the most critical tasks pertinent to the data context. This ensures that users can quickly and effortlessly engage with the most vital aspects of the data, enhancing their overall experience.

Recognizing that users come with varied backgrounds, experiences, and preferences, designers should consider offering customization options. By tailoring the interaction-visualization pairings to specific user groups, the system can become more adaptive and user-centric. One way to harness the design space's flexibility is by providing users with choices. Instead of locking them into a fixed set of interactions, customization can be offered through a menu of options. For instance, if a user is tasked with switching visualization types, they could be presented with a choice at system configuration: would they prefer to ``rotate'' or ``shake'' the tangible cube? Alternatively, the design space can act as a ``dictionary'' for the designers or users to build their own rule handbooks. Such empowerment not only enhances user agency but also ensures that the system remains adaptable to diverse user preferences and needs.

\subsection{Generalizability of the Design Space}

To demonstrate the design space's versatility, we explored its application through two complementary design concepts that analyze demographic and weather data. These two familiar, everyday analytical contexts illustrate the potential of our design space to accommodate a wide spectrum of spatio-temporal data.

\begin{figure*}[ht]
  \centering
  \includegraphics[width=\textwidth]{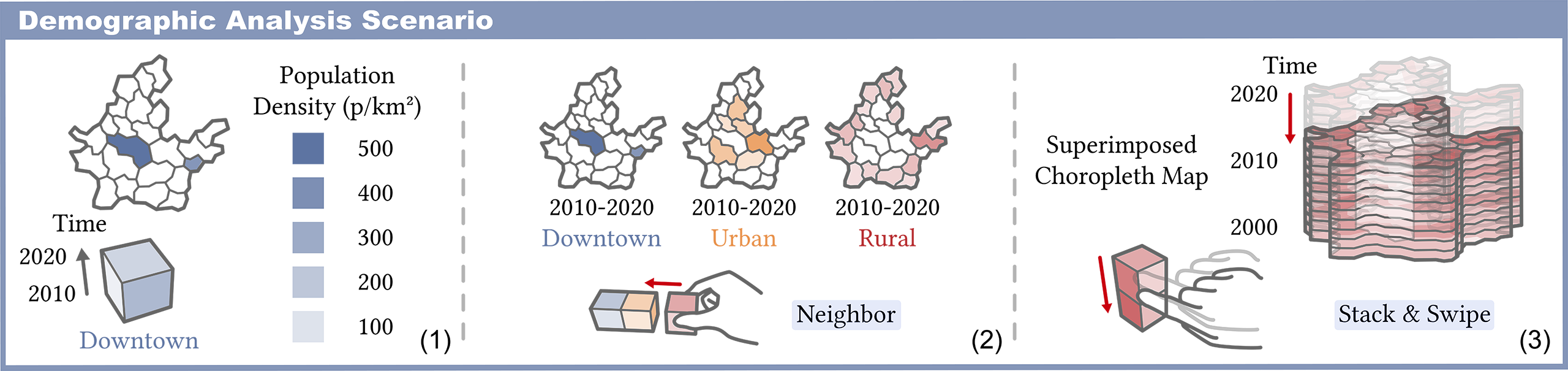}
  \caption{Illustration of the demographic analysis scenario: (1) A space-time cube illustrates the population density of the downtown region. (2) Neighboring three cubes to compare their population density through small multiples. (3) Stacking the cubes generates a superimposed choropleth map, allowing users to swipe along the vertical side to select a specific time range.}
  \Description{A three-sectioned image demonstrating the use of the demographic analysis scenario. The left section shows a map and a cube labelled with 2010-2020 on the vertical edge and Downtown on its horizontal edge. Areas on the map is highlighted with blue. A legend on the right shows different 5 different shades of blue, corresponding to population density of 100-500 in p/km^2. The highlighted area on the map is around the same color as 400-500 p/cm^2. The middle section shows a hand neighboring one cube with two other cubes that are already on the surface. The three cubes are colored blue, orange and red. Above these cubes there are three maps arranged side by side, also colored in blue, orange and red. The blue one represents downtown area from 2010-2020, the orange one represents urban areas from 2010-2020, the red one represents rural areas from 2010-2020. On the right section, a hand swipes down on the vertical axis of a stack of red cubes. This shows a superimposed choropleth map that os gets animated across time periods.}
  \label{fig:demographic}
\end{figure*}

\vspace*{0.5em}
\begin{description}[nosep,leftmargin=1.5em,labelindent=0em,leftmargin=!,labelindent=!,itemindent=!,font=\normalfont\itshape]%
\item[\textbf{Demographic Analysis Scenario:}]
Demographic change is a crucial aspect of sociological studies that focus on population dynamics, migration patterns, and urban development. Such spatio-temporal data provides insights into social shifts and the formation of new population centers. In this scenario (\autoref{fig:demographic}), sociologists can use our tangible design space to immersively explore these demographic changes. The design employs choropleth maps within a space-time cube to visualize changes in population density over time. Each cube represents the population of a specific region over a period of time.

\hspace*{1em}
For this design concept, selecting individual cubes allows users to view detailed demographic data. Placing cubes next to each other reveals comparative population densities through small multiples. Stacking cubes combines data from various regions or time periods, forming a superimposed choropleth map. Additionally, users can swipe along the vertical sides of the stack to choose a specific time range, enabling them to animate and observe demographic changes over time.

\hspace*{1em}
In this example, we highlight how our design space extends beyond the initial prototype to accommodate distinct spatio-temporal data scenarios for specific research needs. Moreover, we demonstrate that the design space's interactions are not confined to one type of visual representation, such as bar charts. It readily adapts to thematic maps and other visual forms. Designers have the freedom to select visual forms that best suit their research and analytical objectives.

\begin{figure*}[ht]
  \centering
  \includegraphics[width=\textwidth]{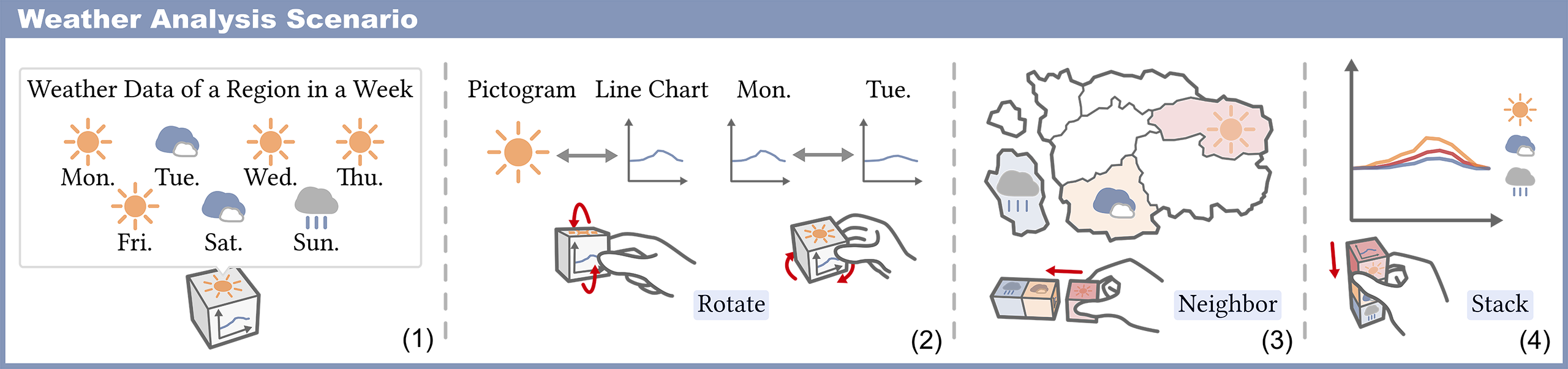}
  \caption{Weather analysis example: (1) The space-time cube represents the weather data of a region in a week. (2) Rotating a cube around the $x$-axis allows users to switch visualization types between categorical weather pictograms and temperature line charts, while rotation around the $y$-axis switches between days in the week. (3) Assembling the cubes creates a synoptic view of regional weather conditions. (4) Stacking the cubes superimposes the line charts for trend comparisons.}
  \Description{A four-sectioned image demonstrating the weather analysis scenario. The first section shows pictograms of weather data in a week, from Mon. To Sun. The pictograms are sunny, cloudy, sunny, sunny, sunny, cloudy and rainy. These pictograms point out from the top surface of the cube with a sun icon on it. On the vertical surface of the cube there is a line-chart icon. In the second section, the cube is held by a hand and rotated in the vertical and horizontal direction. When the cube is rotated vertically around the horizontal axis, the chart switches from a pictogram to a line chart; when the cube is rotated horizontally around the vertical axis, the line chart switches from Mon. To Tue. In the third panel, a hand is neighboring a red cube with a blue and an orange cube that are already on the surface. Above them a synoptic view of these three regions is shown in the map, overlaid with their color and the weather pictograms. In the fourth panel, a hand stacks the red cube on top of the orange and blue cubes. The visualization above shows the superimposed line chart, each line in the same color as the cubes.}
  \label{fig:weather}
\end{figure*}

\hspace*{0.5em}
\item[\textbf{Weather Analysis Scenario:}]
Understanding patterns of weather over time and across locations is essential for meteorologists. This spatio-temporal data includes a blend of quantitative measurements like temperature and precipitation, and categorical conditions like sunny, cloudy, or rainy weather.

\hspace*{1em}
Each tangible cube in this design represents the weather condition of a specified region (e.g. a city) over a week. The visualization employs small multiples of weather charts, combining line charts for quantitative weather data with pictograms for categorical weather conditions. Rotating a cube around its horizontal axis (flipping the cube over its horizontal axis) switches between temperature line graphs and weather condition pictograms. On the other hand, spinning the cube around its vertical axis switches the visualization between days in a week. Placing cubes side-by-side allows for a comparative synoptic view of regional weather on a map. Stacking the cubes superimposes the line charts for trend comparisons. 

\hspace*{1em}
This scenario demonstrates that more than one type of visualization can be incorporated to support the multilayered exploration of data. Taking advantage of the interaction mappings, users can fluidly switch between different views of the multilayered spatio-temporal weather measurements. This flexibility is especially valuable and can be generalized for exploring other multi-dimensional spatio-temporal data.
\end{description}

\vspace{0.5em}
\noindent
Collectively, these two cases that we presented demonstrate how the design space can integrate different data types and visual encodings for different scenarios than the prototype. Furthermore, a key point to note is that although a cube in our main example represents a specific data value, like health spending over a decade, its design inherently offers more versatility. Theoretically, each of the cube's six faces can represent different data dimensions. These dimensions could range from quantitative aspects, such as age or income, to categorical variables like gender or educational level. As for interaction, these cubes can be arranged side by side to compare one dimension, and they can be rotated to compare another. The ease of combination and manipulation empowers users to explore multidimensional data from diverse perspectives.

\subsection{Material Considerations}
The tangible experience in MR environments can be significantly influenced by the choice of materials. Different materials not only offer varied tactile feedback and interaction dynamics, but can also shape the affordability, durability and ease of manufacturing of the system. Therefore, while our prototype utilized magnetic plastic cubes, there are many additional material options with unique benefits for exploratory experimentation.

\vspace{5pt}
\begin{description}[nosep,leftmargin=1.5em,labelindent=0em,leftmargin=!,labelindent=!,itemindent=!,font=\normalfont\itshape]%
\item[\textbf{Soft Touch:}] Softer materials like silicone or foam, introduce a distinct tactile experience that can markedly deviate from rigid materials. Such a combination of soft materials and computational hardwares has been investigated by Fernaeus et al. \cite{fernaeus2012soft} by the concept of ``soft hardware''. They emphasized the transformative potential of soft materials, such as textiles and embroidery in interaction designs. They suggested that soft shapes can afford distinct interactions that bridge the computational and physical experiences.

\hspace*{1em}
Beyond just the tactile experience, soft-bodied materials, owing to their superior elastic properties, open up avenues for a broader spectrum of interactions as well. For instance, they can accommodate interactions like squeezing, bouncing or wiggling, which might not be feasible with rigid body materials. Li et al. \cite{li2015soft} explored this potential with a soft-bodied jumping robot, leveraging the robot's elasticity to achieve highly dynamic motions. Such innovations hint at the expansive possibilities soft materials can bring to interactive designs.

\hspace*{1em}
Furthermore, the gentle touch and novelty of these materials can evoke comfort to enhance user engagement. The playful nature of soft materials has been investigated by interactive design researchers. U\u{g}ur Yavuz et al.'s work of \textit{Design for Playfulness with Interactive Soft Materials} \cite{playfulness2021soft} suggests that soft materials not only enhance the tactile experience but also offer users a richer and more versatile engagement with the data.
\end{description}

\vspace{5pt}
\begin{description}[nosep,leftmargin=1.5em,labelindent=0em,leftmargin=!,labelindent=!,itemindent=!,font=\normalfont\itshape]%
\item[\textbf{Textured Surfaces:}] In our prototype design, we used tangible cubes with smooth surfaces. However, the introduction of varied textures, including ridged, patterned or other tactile variations, can provide additional interactive merits. By incorporating unique textures on different cubes or even on separate faces of the same cube, we can introduce an additional haptic-based data encoding channel. This tactile differentiation can serve as a neat method for distinguishing individual cubes or categorizing cubes representing different data clusters. Using such tactile differentiation can be helpful in mitigating cognitive load, especially when scaling up the system with a larger quantity of cubes to accommodate more complex datasets.

\hspace*{1em}
Textures can also be considered for enhancing accessibility and inclusivity in the tangible cube designs for different assistive needs. Textures have long been employed in tangible interfaces such as Fan and Antle's design \cite{Fan2015Tactile}, where they explored the use of texture cues in tangible tabletops to support alphabetic learning for dyslexic children. Textures can also facilitate users from the visually impaired community. For instance, textures derived from braille can be seamlessly integrated into the designs, supporting the manipulation of the tangible cubes. These considerations can enable a broader user base to experience the benefits of the system. 

\begin{edited}
\hspace*{1em}
In the context of our design space, we recognize the potential of textures to offer an additional haptic-based data encoding channel. However, we also acknowledge the challenges posed by perceptual variability and the necessity for empirical research to inform the effective use of textures in tangible data representations.
Recent studies, such as Xu et al. \cite{xu2023lets}, emphasize the need to consider perceptual variability in these tactile data encodings. This is particularly relevant in the context of designing textured tangible cubes. Exploring how perceptually relevant tactile surfaces can be combined with visual feedback will enable us to refine our design choices and ensure that our material considerations align with the goals of accessibility and effective data communication.
\end{edited}
\end{description}

\begin{figure}
  \centering
  \includegraphics[width=0.49\columnwidth]{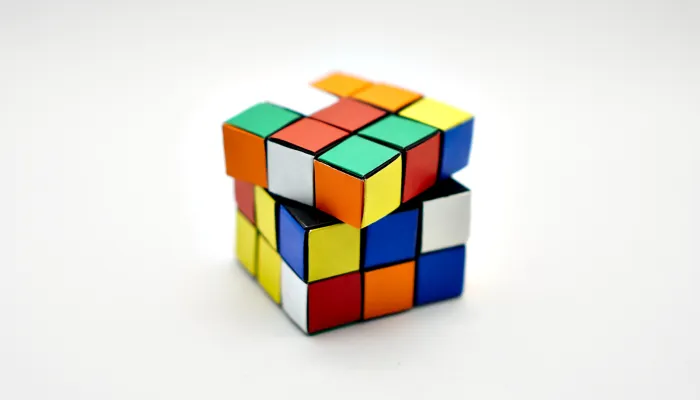}
 \includegraphics[width=0.49\columnwidth]{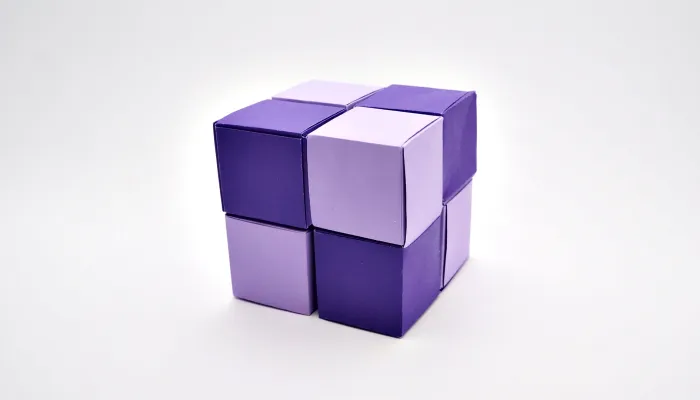}
  \caption{Images of the origami rubik cube and origami infinity cube created by Nakashima.
  \cite{nakashima2019origami}.}
  \Description{Two side-by-side images showcasing intricate origami cube designs. On the left is the 'origami rubik cube', which resembles a Rubik's cube but is crafted entirely from folded paper. On the right is the 'origami infinity cube', a complex paper folding design that appears to have a continuous, looping structure. Both creations are the work of Nakashima and are displayed against a neutral background.}
  \label{fig:origamicubes}
\end{figure}

\vspace{5pt}
\begin{description}[nosep,leftmargin=1.5em,labelindent=0em,leftmargin=!,labelindent=!,itemindent=!,font=\normalfont\itshape]%
\item[\textbf{Paper Cubes:}] Paper presents a ubiquitous avenue for constructing tangible cubes. Origami techniques offer a rich repertoire of methods to transform papers into a myriad of intricate cube designs. For instance, paper origami designs such as the \textit{Origami Rubik's Cube} and the \textit{Origami Infinity Cube} by Nakashima \cite{nakashima2019origami} showcase the feasibility and versatility of paper as a potential medium for creating tangible cubes (\autoref{fig:origamicubes}). 

\hspace*{1em}
Furthermore, paper's universal availability and adaptability make it particularly advantageous for educational contexts and rapid prototyping scenarios. In such settings, users can engage in hands-on activities, customizing or crafting their tangible cubes on-the-spot. This in-situ prototyping approach fosters creativity, allowing users to iterate and refine their designs in real time. By using paper, educators and workshop facilitators can provide a tangible, interactive experience without incurring significant costs.

\end{description}

\vspace{1em}
\noindent
In summary, the material choices we discussed here are provided as inspirations and merely a glimpse into the vast possibilities for crafting innovative tangible cubes. 

\subsection{Implementation Techniques}
Implementing tangible cubes, particularly when multiple cubes are involved, requires precise and real-time tracking of each cube's position and orientation. The method chosen to track these cubes is hence a critical aspect of the technical implementation of such tangible cube systems in MR. While there are various technical approaches available for tracking individual cubes, we outline a few prevalent strategies that have been adopted by relevant tangible interaction systems.

\vspace{5pt}
\begin{description}[nosep,leftmargin=1.5em,labelindent=0em,leftmargin=!,labelindent=!,itemindent=!,font=\normalfont\itshape]%
\item[\textbf{Optical Tracking}]techniques employ cameras or infrared sensors to detect the position and orientation of tangible objects. This approach is beneficial due to its non-intrusiveness and ability to simultaneously track multiple objects. However, it requires a clear line of sight between the camera and the cubes, and can be significantly impacted by occlusion and distance. Different variants of optical tracking techniques can be used, depending on the specific precision and implementation needs. Here we describe two subset techniques under this category of optical tracking:

\begin{itemize}
\item \textit{\textbf{Computer Vision with Markers or Feature Points}}: Various computer vision algorithms exist for object tracking. These algorithms typically rely on the detection and recognition of either feature points on image-based targets or human-made markers that are attached to the object. Images with rich feature patterns or markers with varied structures and designs (e.g., QR codes, Hiro markers and ArUco markers) can be attached to the surfaces of each tangible cube to serve as targets for detection. These targets can then be effectively captured by modern computer vision algorithms in real-time interactions. For instance, Tong et al. \cite{tong2023exploring} implemented feature point identification on paper sheets using Vuforia to track their position and orientation for interacting with printed data visualizations. Jansen et al. \cite{jansen2020share} employed computer vision techniques through cameras on HMDs to identify Hiro markers for accurately calibrating projections onto tangible surfaces in their AR system. 

\item \textit{\textbf{Reflective Sensors and Motion Capture}:} Reflective sensors detect light reflected off markers placed on objects, while motion capture uses specialized cameras for triangulating this data and pinpointing the object's location and orientation. For larger cube entities or environments where precision is crucial, this combination offers high-fidelity tracking capabilities. For instance, Ens. et al. \cite{ens2020uplift} employed this method in their ``Uplift'' system, using Vicon\footnote{https://www.vicon.com/} to track widgets for supporting ``casual collaborative visual analytics''.
\end{itemize}
\end{description}

\vspace{5pt}
\begin{description}[nosep,leftmargin=1.5em,labelindent=0em,leftmargin=!,labelindent=!,itemindent=!,font=\normalfont\itshape]%
\item[\textbf{Inertial Tracking}]relies on sensors that measure physical properties like acceleration and rotations. A common category of sensors used is inertia measurement units (IMUs). These compact sensors can be embedded within the tangible cubes to detect changes in movement and orientation based on accelerometers, gyroscopes, and sometimes magnetometers. As an example, Kaimoto et al. \cite{kaimoto2022sketched} implemented an AR sketching interactive system using small tangible user interfaces containing IMUs. Depending on specific contexts, IMUs of varying sizes and precision levels can be chosen. However, they may require periodic recalibration to counteract drifts. 
\end{description}

\vspace{5pt}
\begin{description}[nosep,leftmargin=1.5em,labelindent=0em,leftmargin=!,labelindent=!,itemindent=!,font=\normalfont\itshape]%
\item[\textbf{Wireless Communication}]enables data transfer without the need for physical connections. RFID (Radio Frequency Identification) and NFC (Near Field Communication) tags are specific examples which use radio frequencies to identify, track and store information on objects in tangible systems. For example, Hosokawa et al. \cite{hosokawa2008tangible} used RFID technology in a system that enabled users to design their own houses. Lee et al. \cite{lee2022nfcstack} presented a physical building block system based on NFC. When embedded within the cubes, these tags not only facilitate tracking but also allow for data storage, making them particularly advantageous in scenarios where individual cubes might need to retain specific data or settings. However, while these wireless methods offer convenience, they can face challenges in conditions like signal interference or range limitations.
\end{description}

\begin{edited}
\subsection{Balancing Immersion with Accessibility}
While MR environments offer unparalleled immersion, deployment beyond research contexts obliges deliberation on accessibility. Educational and cultural settings prioritize inclusivity yet operate on restricted budgets-requiring prudent design decisions balancing experience with practicality.

Material choices significantly impact both facets. Origami paper techniques enable serviceable cubes nearly cost-free, lending well to hands-on classrooms. Implementations should also consider trade-offs; hardware solutions provide responsiveness yet adversely impact expenses. Hybrids like computer vision tracking with wireless data transmission optimize both accuracy and affordability.

Display systems also warrant accessibility-driven reimagination. Large-format CAVEs visualize multifaceted dynamics beautifully yet remain prohibitively expensive at scale. Instead, existing in-situ projection infrastructure at museums alongside visitors’ own mobile devices could host these tangible interactions at fractions of the cost. While no solution perfectly reconciles immersion with accessibility, perceiving challenges as design opportunities encourages us toward equitable, captivating data exploration for diverse contexts.

In summary, designing for cost and accessibility involves careful material selection, implementation techniques, and display choices, ensuring that these MR tangible interactive tools are available to a wider audience.

\end{edited}

\section{Conclusion}
Our journey through the design space of tangible mixed-reality cubes has provided valuable insights into the potential of such interfaces for data visualization. By reviewing relevant literature and elucidating user insights through an ideation workshop, we derived a comprehensive design space of interactions and visualizations based on tangible cubes tailored for mixed-reality environments. We created a proof-of-concept prototype and evaluated its performance through a qualitative user study. From the feedback, we identified key areas for further exploration, including scalability challenges, material choices, and the balance between flexibility and customization. These discussions not only highlight the design space's potential but also offer practical guidance for future designers.

As we look forward, there are many opportunities to expand on this work. Exploring how tangible cubes can handle more complex datasets, experimenting with different cube materials, and integrating new tracking methods are promising next steps. With continued research and development, the design space for tangible mixed-reality cubes can lead to even more intuitive and immersive user experiences.

\begin{acks}
This work was supported by the National Natural Science Foundation of China 62272396.
\end{acks}

\bibliographystyle{ACM-Reference-Format}
\bibliography{reference}

\newpage
\appendix

\section{List of Data Tasks and Questions Used in the Workshop}
\label{app: appendix1}

\textbf{Explore}
\begin{enumerate}
    \item Explore the health spending data of all countries from 2000-2020
    \item Explore the health spending data of all countries from 2000-2010
    \item Explore the health spending data of all countries from 2000-2005
\end{enumerate}

\noindent
\textbf{Combination}
\begin{enumerate}
    \item Combine the health spending of China and Australia
    \item Combine the health spending of China, Australia, US
\end{enumerate}

\noindent
\textbf{Difference}
\begin{enumerate}
    \item Find the difference in health spending between China and Australia
    \item Find the difference in health spending between China, Australia, US
\end{enumerate}

\noindent
\textbf{Snapshot}
\begin{enumerate}
    \item Generate a 2D snapshot for all countries over time
\end{enumerate}

\noindent
\textbf{Annotate}
\begin{enumerate}
    \item Annotate / highlight the largest value from the previous step
\end{enumerate}

\noindent
\textbf{Re-Scale}
\begin{enumerate}
    \item How would you re-scale the data scale each cube represents?
\end{enumerate}

\section{User Study Interview Questions}

\label{app:appendix2}
\textbf{Size and Multiplicity of the Tangible Cubes}
\begin{enumerate}
    \item How comfortable were you with the sizes of the cubes: Were they easy to handle and manipulate?

    \item What are your thoughts on the multiplicity of the cubes: Did having multiple cubes help or hinder your ability to perform the tasks?

    \item What are your thoughts on combining smaller cubes into larger ones: Did you find it intuitive or challenging?
\end{enumerate}

\noindent
\textbf{Interaction Space}
\begin{enumerate}
    \item Did you find the interaction tasks easy to perform with the tangible cubes and the associated actions?
    
    \item Which task was the easiest, and which was the hardest for you, and why?
\end{enumerate}

\noindent
\textbf{Visualization Space}
\begin{enumerate}
    \item Were you comfortable with the way the visualization is presented, both embodied on the individual cubes and together in the designated visualization area?
    \item Was it easy to understand the connection between the tangible cube manipulations and the resulting visualizations?
\end{enumerate}

\noindent
\textbf{Holistic Questions}
\begin{enumerate}
    \item How effective do you think tangible cube interactions are in visualizing and exploring complex datasets?
    \item What are the potential advantages and disadvantages you see in this system?
    \item Can you imagine using such a system  in a real-world context, such as in a research setting or an educational environment? What changes or improvements would you suggest?
\end{enumerate}

\end{document}

%% file: table/cube_taxonomy.tex
\newcommand{\cmark}{\scalebox{1.2}{$\blacksquare$}} 
\newcommand{\xmark}{\scalebox{1.2}{$\square$}} 

\begin{table}[htbp]
\centering

\setlength{\tabcolsep}{1.4pt} 
\caption{Summary of the use of tangible cubes in the literature from four design aspects: size, interaction space, visualization space and multiplicity.}
\label{tab:summary}
\tiny
\begin{tabular}{lll c cccccc cc cccccc c cccccc}

{\fontsize{6}{9}\selectfont
\textbf{Size}} & & {\fontsize{6}{9}\selectfont \textbf{Related Work}} & & \multicolumn{6}{c}{{\fontsize{6}{9}\selectfont\textbf{Interaction Space}}} & & & \multicolumn{6}{c}{{\fontsize{6}{9}\selectfont\textbf{Visualization Space}}} & & \multicolumn{6}{c}{{\fontsize{6}{9}\selectfont\textbf{Multiplicity}}}\\
\toprule
& & & & & \rotatebox{90}{{\fontsize{6}{9}\selectfont\textbf{Orientation}}} & \rotatebox{90}{{\fontsize{6}{9}\selectfont\textbf{Translation}}} & \rotatebox{90}{{\fontsize{6}{9}\selectfont\textbf{Combination}}} & \rotatebox{90}{{\fontsize{6}{9}\selectfont\textbf{Surface}}} & \rotatebox{90}{{\fontsize{6}{9}\selectfont\textbf{Transformation}}} & & & \rotatebox{90}{{\fontsize{6}{9}\selectfont\textbf{Overlay}}} & \rotatebox{90}{{\fontsize{6}{9}\selectfont\textbf{Above}}} & \rotatebox{90}{{\fontsize{6}{9}\selectfont\textbf{Side}}} & \rotatebox{90}{{\fontsize{6}{9}\selectfont\textbf{Display}}} & \rotatebox{90}{{\fontsize{6}{9}\selectfont\textbf{Inside}}} & \rotatebox{90}{{\fontsize{6}{9}\selectfont\textbf{Around}}} & & & & \rotatebox{90}{{\fontsize{6}{9}\selectfont\textbf{Multiple}}} & \rotatebox{90}{{\fontsize{6}{9}\selectfont\textbf{Single}}} & & \\
\cmidrule(r){6-11} \cmidrule(r){13-19} \cmidrule(r){22-24}


Small & & Juan10 \cite{juan2010tangible} & & & \cmark & \xmark & \xmark & \xmark & \xmark & & & \cmark & \xmark & \xmark & \xmark & \xmark & \xmark & & & & \cmark & \xmark & & \\

Small & & Grasset07 \cite{grasset2007mixed} & &  & \xmark & \cmark & \xmark & \xmark & \xmark & & & \xmark & \cmark & \xmark & \xmark & \xmark & \xmark & & & & \xmark & \cmark & & \\

Small & & Gong19 \cite{gong2019grey} & & & \cmark & \xmark & \xmark & \xmark & \xmark & & & \xmark & \xmark & \cmark & \xmark & \xmark & \xmark & & & & \cmark & \xmark & & \\

Small & & Wittkopf06 \cite{wittkopfi3} & & & \xmark & \cmark & \xmark & \xmark & \xmark & & & \xmark & \xmark & \cmark & \xmark & \xmark & \xmark & & & & \xmark & \cmark & & \\

Small & & Ha10 \cite{ha2010empirical} & & & \cmark & \cmark & \xmark & \cmark & \xmark & & & \xmark & \xmark & \cmark & \xmark & \xmark & \xmark & & & & \xmark & \cmark & & \\

Small & & Ma20 \cite{ma2020mixed} & & & \cmark & \xmark & \xmark & \xmark & \xmark & & & \xmark & \xmark & \xmark & \cmark & \xmark & \xmark & & & & \cmark & \xmark & & \\

Small & & Langner14 \cite{langner2014cubequery} & & & \xmark & \xmark & \cmark & \xmark & \xmark & & & \xmark & \xmark & \xmark & \cmark & \xmark & \xmark & & & & \cmark & \xmark & & \\

\midrule


Medium & & Bozgeyiki21 \cite{bozgeyikli2021evaluating} & & & \xmark & \cmark & \xmark & \xmark & \xmark & & & \cmark & \xmark & \xmark & \xmark & \xmark & \xmark & & & & \xmark & \cmark & & \\

Medium & & Grandhi19 \cite{grandhi2019playgami} & & & \xmark & \xmark & \xmark & \xmark & \cmark & & & \cmark & \xmark & \xmark & \xmark & \xmark & \xmark & & & & \xmark & \cmark & & \\

Medium & & Olim20 \cite{ha2010empirical} & & & \cmark & \xmark & \xmark & \xmark & \xmark & & & \cmark & \cmark & \xmark & \xmark & \xmark & \xmark & & & & \xmark & \cmark & & \\

Medium & & Song19 \cite{song2019turtlego} & & & \cmark & \xmark & \cmark & \xmark & \xmark & & & \cmark & \cmark & \xmark & \xmark & \xmark & \xmark & & & & \cmark & \xmark & & \\

Medium & & Bergig11 \cite{bergig2011out} & & & \cmark & \xmark & \xmark & \cmark & \cmark & & & \cmark & \cmark & \xmark & \xmark & \xmark & \xmark & & & & \cmark & \xmark & & \\

Medium & & Hoe19 \cite{hoe2019using} & & & \cmark & \xmark & \xmark & \xmark & \xmark & & & \xmark & \xmark & \cmark & \xmark & \xmark & \xmark & & & & \xmark & \cmark & & \\

Medium &  & Zhou04 \cite{zhou2004magic} & & & \xmark & \cmark & \xmark & \xmark & \xmark & & & \xmark & \xmark & \cmark & \xmark & \xmark & \xmark & & & & \xmark & \cmark & & \\

Medium &  & Qi05 \cite{qi2005tangible} & & & \cmark & \xmark & \xmark & \xmark & \xmark & & & \xmark & \xmark & \xmark & \cmark & \xmark & \xmark & & & & \xmark & \cmark & & \\

Medium &  & Kruzynski08 \cite{kruszynski2008tangible} & & & \cmark & \xmark & \xmark & \xmark & \xmark & & & \xmark & \xmark & \xmark & \cmark & \xmark & \xmark & & & & \xmark & \cmark & & \\

Medium &  & Salem07 \cite{salem2007intercube} & & & \cmark & \xmark & \xmark & \xmark & \xmark & & & \xmark & \xmark & \xmark & \cmark & \xmark & \xmark & & & & \xmark & \cmark & & \\

Medium &  & Lee20 \cite{lee2020using} & & & \cmark & \xmark & \xmark & \xmark & \xmark & & & \xmark & \xmark & \xmark & \cmark & \xmark & \xmark & & & & \xmark & \cmark & & \\

Medium &  & Chakraborty14 \cite{chakraborty2014captive} & & & \cmark & \xmark & \xmark & \xmark & \xmark & & & \xmark & \xmark & \xmark & \xmark & \cmark & \xmark & & & & \xmark & \cmark & & \\

Medium &  & Issartel16 \cite{issartel2016tangible} & & & \xmark & \cmark & \xmark & \xmark & \xmark & & & \xmark & \xmark & \xmark & \xmark & \cmark & \xmark & & & & \xmark & \cmark & & \\

Medium &  & Kim20 \cite{kim2020inside} & & & \cmark & \cmark & \xmark & \xmark & \xmark & & & \xmark & \xmark & \xmark & \xmark & \cmark & \xmark & & & & \xmark & \cmark & & \\

Medium &  & Lee11 \cite{lee2011two} & & & \cmark & \cmark & \cmark & \xmark & \xmark & & & \xmark & \xmark & \xmark & \xmark & \cmark & \xmark & & & & \cmark & \xmark & & \\

Medium &  & Cordeil17 \cite{cordeil2017design} & & & \cmark & \xmark & \xmark & \cmark & \xmark & & & \xmark & \xmark & \xmark & \xmark & \cmark & \xmark & & & & \xmark & \cmark & & \\

Medium &  & Lee10 \cite{lee2010tangible} & & & \cmark & \xmark & \xmark & \xmark & \cmark & & & \xmark & \xmark & \xmark & \xmark & \xmark & \cmark & & & & \cmark & \xmark & & \\

\midrule


Large &  & DeLariviere08 \cite{de2008cubtile} & & & \xmark & \xmark & \xmark & \cmark & \xmark & & & \xmark & \xmark & \xmark & \cmark & \xmark & \xmark & & & & \xmark & \cmark & & \\

Large &  & Rinott13 \cite{rinott2013cubes} & & & \cmark & \xmark & \cmark & \cmark & \xmark & & & \xmark & \xmark & \xmark & \xmark & \cmark & \xmark & & & & \xmark & \cmark & & \\

\midrule


Various &  & Cleto20 \cite{cleto2020Code} & & & \xmark & \xmark & \cmark & \xmark & \xmark & & & \xmark & \cmark & \xmark & \xmark & \xmark & \xmark & & & & \cmark & \xmark & & \\

Various &  & Hsu22 \cite{hsu2022based} & & & \xmark & \xmark & \xmark & \xmark & \cmark & & & \xmark & \xmark & \cmark & \xmark & \xmark & \xmark & & & & \cmark & \xmark & & \\

\bottomrule

\end{tabular}
\vspace{-5ex}
\end{table}

%% file: table/gestures.tex
\begin{table}[htb]
    \caption{Summary of tangible cube interaction gestures, categorized by their modality. The table presents single-touch gestures, multi-touch gestures, surface trajectories, and hover gestures, each accompanied by a visual representation and a brief description.}
    \centering
    \begin{tabular}{|c|c|c|}
    \hline
        \begin{minipage}{0.49\linewidth} 
            \vspace{5pt}
            \noindent
            \textbf{\textit{Single-Touch Gestures}} are basic interactions involving one point of contact with the tangible cube, typically with one finger.
        
            \setlength\intextsep{0pt}
            \setlength\columnsep{0pt}
            \begin{wrapfigure}{l}{.12\columnwidth}
              \vspace*{\fill}
              \includegraphics[height=1.6\baselineskip]{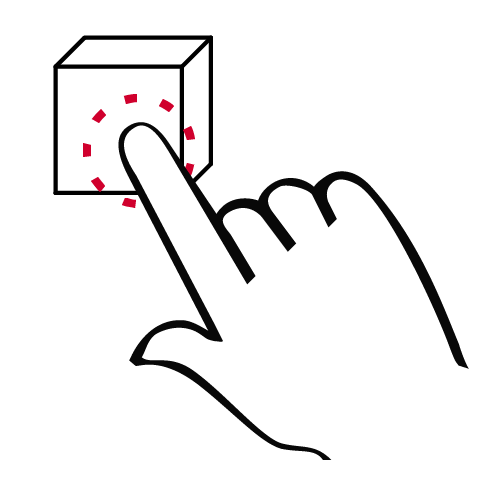}
              \vspace*{\fill}
            \end{wrapfigure}
            \noindent
            \underline{\textit{Tap}}: quickly touching and releasing the surface of a cube with one finger.
            
            \setlength\intextsep{0pt}
            \setlength\columnsep{0pt}
            \begin{wrapfigure}{l}{.12\columnwidth}
              \vspace*{\fill}
              \includegraphics[height=1.6\baselineskip]{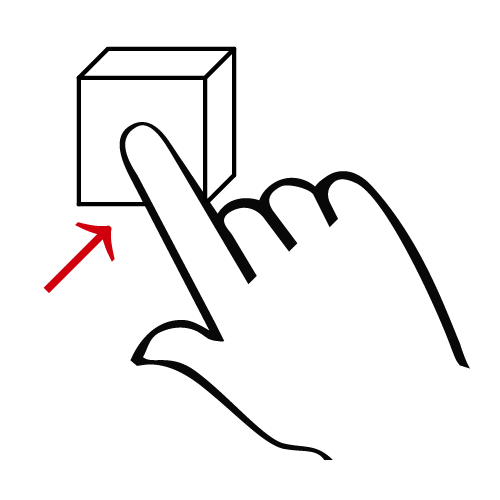}
              \vspace*{\fill}
            \end{wrapfigure}
            \noindent
            \underline{\textit{Press}}: applying varying pressure on the cube using one finger for activation. 
            
            \setlength\intextsep{0pt}
            \setlength\columnsep{0pt}
            \begin{wrapfigure}{l}{.12\columnwidth}
              \vspace*{\fill}
              \includegraphics[height=1.4\baselineskip]{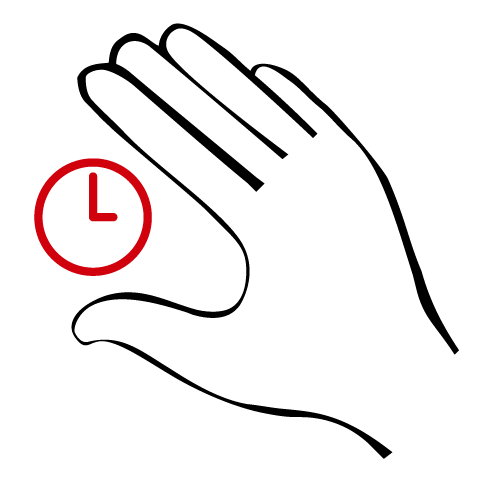}
              \vspace*{\fill}
            \end{wrapfigure}
            \noindent
            \underline{\textit{Hold}}: maintaining contact for an extended period as a continuous state.
            \vspace{5pt}
        \end{minipage} 
    & 
        \begin{minipage}{0.49\linewidth} 
            \vspace{5pt}
            \noindent
            \textbf{\textit{Multi-Touch Gestures}} use multiple points of contact or touch the cube multiple times, performed using one or more fingers.
    
            \setlength\intextsep{0pt}
            \setlength\columnsep{0pt}
            \begin{wrapfigure}{l}{.12\columnwidth}
              \vspace*{\fill}
              \includegraphics[height=1.6\baselineskip]{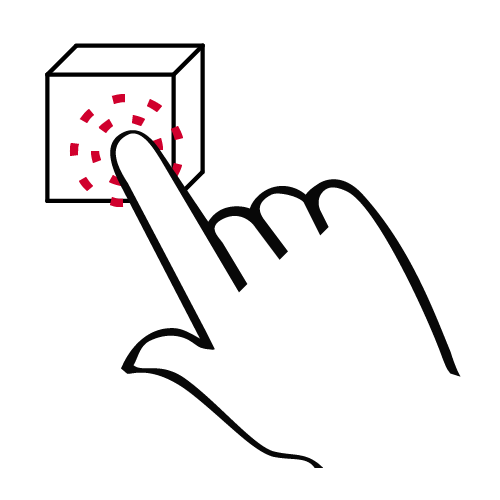}
              \vspace*{\fill}
            \end{wrapfigure}
            \noindent
            \underline{\textit{Double Tap}}: tapping the surface of a cube twice in quick succession with one finger.
            
            \setlength\intextsep{0pt}
            \setlength\columnsep{0pt}
            \begin{wrapfigure}{l}{.12\columnwidth}
              \vspace*{\fill}
              \includegraphics[height=1.6\baselineskip]{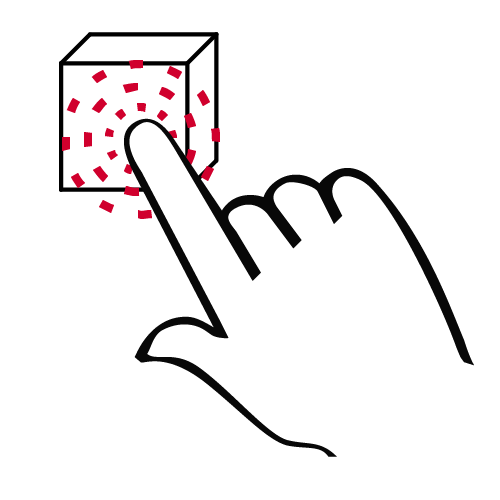}
              \vspace*{\fill}
            \end{wrapfigure}
            \noindent
            \underline{\textit{Triple Tap}}: tapping a cube surface three times in quick succession with one finger.
            
            \setlength\intextsep{0pt}
            \setlength\columnsep{0pt}
            \begin{wrapfigure}{l}{.12\columnwidth}
              \vspace*{\fill}
              \includegraphics[height=1.6\baselineskip]{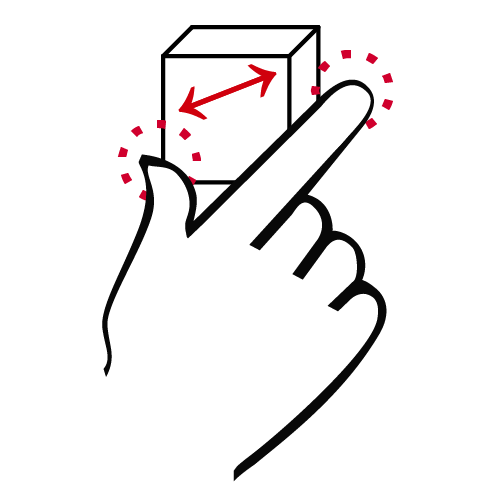}
              \vspace*{\fill}
            \end{wrapfigure}
            \noindent
            \underline{\textit{Pinch}}: establishing two contact points and varying their distance on the cube surface. 
            \vspace{5pt}
        \end{minipage}
    \\ 
        \begin{minipage}{0.49\linewidth} 
            \vspace{5pt}
            \noindent
            \textbf{\textit{Surface Trajectories}} involve surface interactions that use continuous finger movements to create trajectories.

            \setlength\intextsep{0pt}
            \setlength\columnsep{0pt}
            \begin{wrapfigure}{l}{.12\columnwidth}
              \vspace*{\fill}
              \includegraphics[height=1.6\baselineskip]{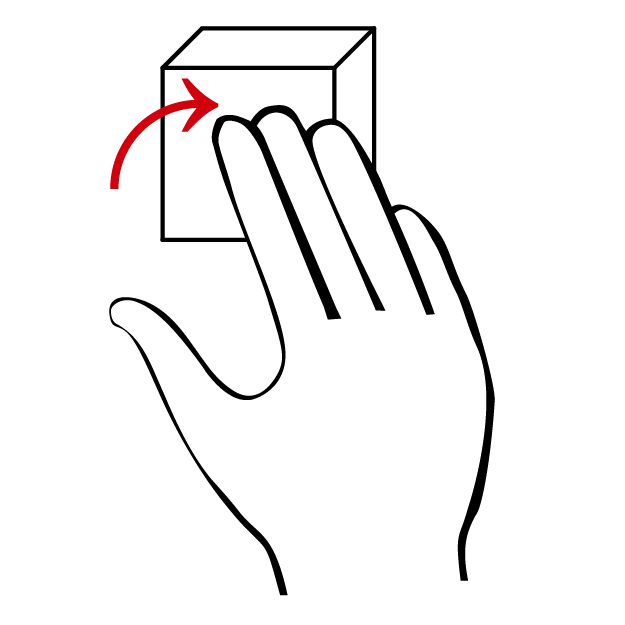}
              \vspace*{\fill}
            \end{wrapfigure}
            \noindent
            \underline{\textit{Swipe}}: establishing a contact point, then moving across the surface to create a linear trajectory with one finger.

            \setlength\intextsep{0pt}
            \setlength\columnsep{0pt}
            \begin{wrapfigure}{l}{.12\columnwidth}
              \vspace*{\fill}
              \includegraphics[height=1.6\baselineskip]{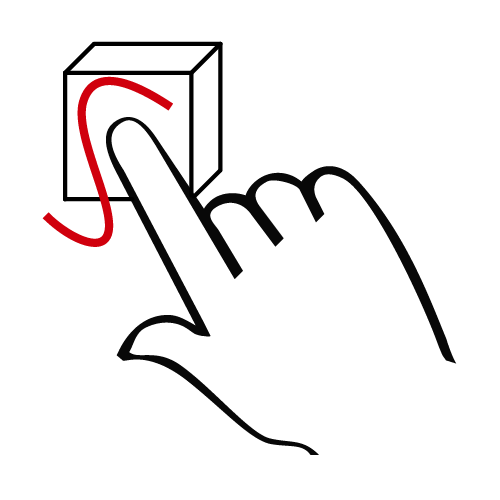}
              \vspace*{\fill}
            \end{wrapfigure}
            \noindent
            \underline{\textit{Path}}: establishing a contact point, then moving to create a linear or curved trajectory that may span multiple cube faces.
            \vspace{5pt}
        \end{minipage}
    & 
        \begin{minipage}{0.49\linewidth} 
            \vspace{5pt}
            \noindent
            \textbf{\textit{Hover Gestures}} refer to positioning fingers or hands on top of a cube without making firm physical contact.

            \setlength\intextsep{0pt}
            \setlength\columnsep{0pt}
            \begin{wrapfigure}{l}{.12\columnwidth}
              \vspace*{\fill}
              \includegraphics[height=1.6\baselineskip]{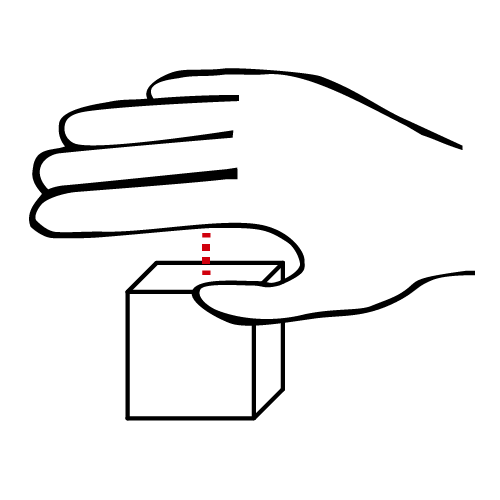}
              \vspace*{\fill}
            \end{wrapfigure}
            \noindent
            \underline{\textit{Open Palm Hover}}: positioning a hand directly above a cube without touching it.

            \setlength\intextsep{0pt}
            \setlength\columnsep{0pt}
            \begin{wrapfigure}{l}{.12\columnwidth}
              \vspace*{\fill}
              \includegraphics[height=1.6\baselineskip]{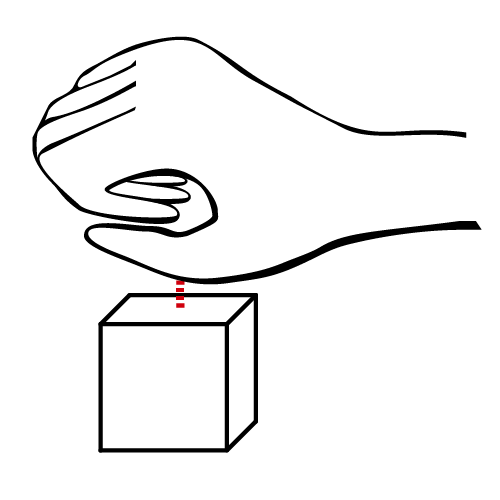}
              \vspace*{\fill}
            \end{wrapfigure}
            \noindent
            \underline{\textit{Closed Fist Hover}}: hovering a closed fist directly above a cube without touching it.
            
            \setlength\intextsep{0pt}
            \setlength\columnsep{0pt}
            \begin{wrapfigure}{l}{.12\columnwidth}
              \vspace*{\fill}
              \includegraphics[height=1.6\baselineskip]{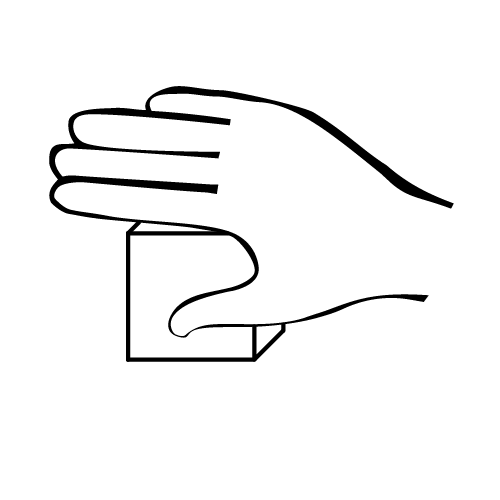}
              \vspace*{\fill}
            \end{wrapfigure}
            \noindent
            \underline{\textit{Cover}}: positioning a hand above a cube, fully or partially occluding it from view. 
            \vspace{5pt}
        \end{minipage}
    \\ 
    \hline
    \end{tabular}
    \label{tab:Gestures}
\end{table}

%% file: table/physical_manipulations.tex
\begin{table}[htb]
    \caption{Summary of physical manipulations with tangible cubes, categorized by the number of cubes involved. The table delineates single-cube manipulations, such as picking up or rotating, and multi-cube manipulations, like stacking or neighboring, each complemented by a visual representation and a description.}
    \centering
    \begin{tabular}{|c|c|c|}
    \hline
        \begin{minipage}{0.49\linewidth} 
            \vspace{5pt}
            \noindent
            \textbf{\textit{Single-Cube Manipulations}} are performed on an individual cube or a set of cubes that have been assembled into a cohesive unit.
        
            \setlength\intextsep{0pt}
            \setlength\columnsep{0pt}
            \begin{wrapfigure}{l}{.12\columnwidth}
              \vspace*{\fill}
              \includegraphics[height=1.6\baselineskip]{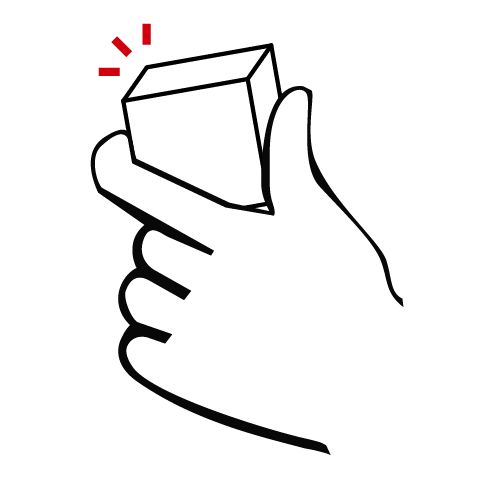}
              \vspace*{\fill}
            \end{wrapfigure}
            \noindent
            \underline{\textit{Pick Up}}: grasping and lifting a cube away from its baseline or resting surface level.
            
            \setlength\intextsep{0pt}
            \setlength\columnsep{0pt}
            \begin{wrapfigure}{l}{.12\columnwidth}
              \vspace*{\fill}
              \includegraphics[height=1.6\baselineskip]{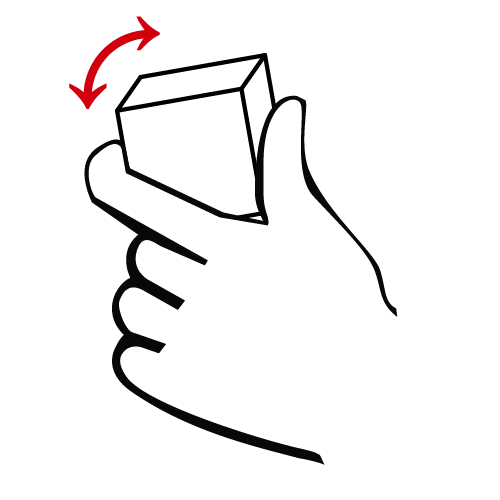}
              \vspace*{\fill}
            \end{wrapfigure}
            \noindent
            \underline{\textit{Rotate}}: adjusting the orientation of a cube by turning it around its central axis.
            
            \setlength\intextsep{0pt}
            \setlength\columnsep{0pt}
            \begin{wrapfigure}{l}{.12\columnwidth}
              \vspace*{\fill}
              \includegraphics[height=1.6\baselineskip]{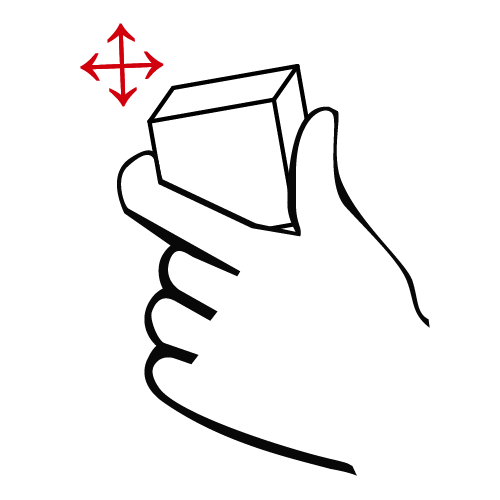}
              \vspace*{\fill}
            \end{wrapfigure}
            \noindent
            \underline{\textit{Translate}}: repositioning a cube on a plane without altering its existing orientation.

            \setlength\intextsep{0pt}
            \setlength\columnsep{0pt}
            \begin{wrapfigure}{l}{.12\columnwidth}
              \vspace*{\fill}
              \includegraphics[height=1.6\baselineskip]{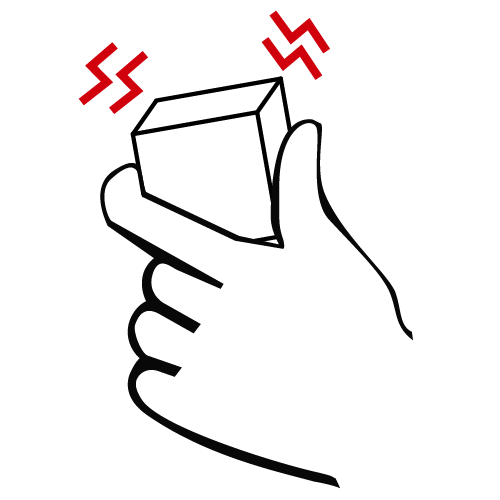}
              \vspace*{\fill}
            \end{wrapfigure}
            \noindent
            \underline{\textit{Shake}}: quickly and dynamically moving a cube up and down or side to side.
            \vspace{5pt}
        \end{minipage} 
    & 
        \begin{minipage}{0.49\linewidth} 
            \vspace{5pt}
            \noindent
            \textbf{\textit{Multi-Cube Manipulations}} are actions performed on multiple cubes or sets of cubes, which are manipulated as separate entities.
    
            \setlength\intextsep{0pt}
            \setlength\columnsep{0pt}
            \begin{wrapfigure}{l}{.12\columnwidth}
              \vspace*{\fill}
              \includegraphics[height=1.6\baselineskip]{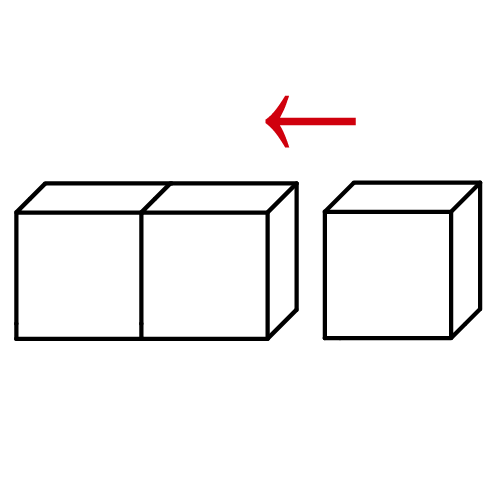}
              \vspace*{\fill}
            \end{wrapfigure}
            \noindent
            \underline{\textit{Neighbor}}: placing two cubes side to side on a plane, forming a horizontal alignment.
            
            \setlength\intextsep{0pt}
            \setlength\columnsep{0pt}
            \begin{wrapfigure}{l}{.12\columnwidth}
              \vspace*{\fill}
              \includegraphics[height=1.6\baselineskip]{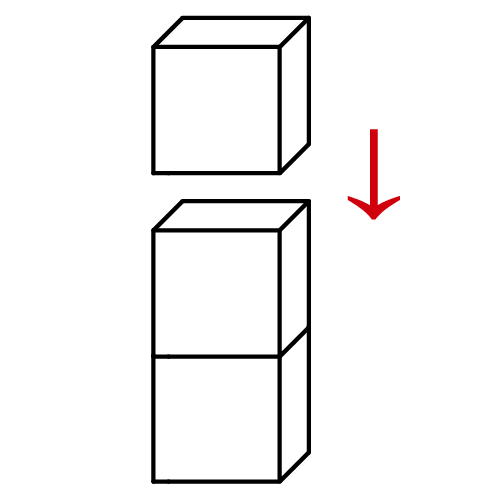}
              \vspace*{\fill}
            \end{wrapfigure}
            \noindent
            \underline{\textit{Stack}}: placing two or more cubes atop one another, forming a vertical column.
            
            \setlength\intextsep{0pt}
            \setlength\columnsep{0pt}
            \begin{wrapfigure}{l}{.12\columnwidth}
              \vspace*{\fill}
              \includegraphics[height=1.4\baselineskip]{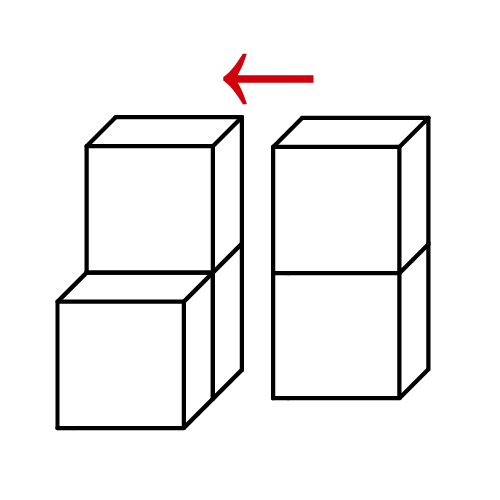}
              \vspace*{\fill}
            \end{wrapfigure}
            \noindent
            \underline{\textit{Assemble}}: engaging three or more cubes to form a larger cohesive structure.

            \setlength\intextsep{0pt}
            \setlength\columnsep{0pt}
            \begin{wrapfigure}{l}{.12\columnwidth}
              \vspace*{\fill}
              \includegraphics[height=1.6\baselineskip]{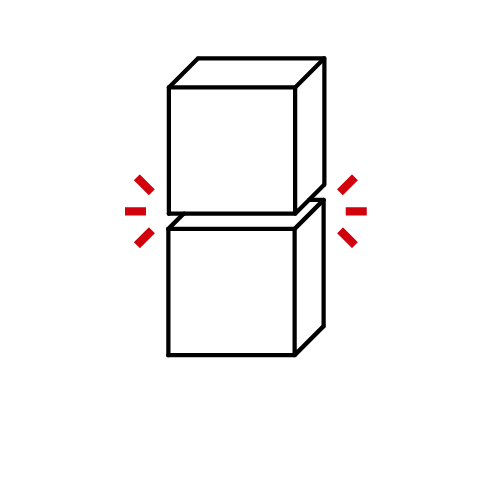}
              \vspace*{\fill}
            \end{wrapfigure}
            \noindent
            \underline{\textit{Collide}}: forcefully moving two cubes towards each other with momentum.
            \vspace{5pt}
        \end{minipage}
    \\ 
    \hline
    \end{tabular}
    \label{tab:Physical Manipulations}
\end{table}

%% file: table/use_design_space.tex
\begin{table}[h]
\centering
\caption{Example set of one-to-one mappings drawn from the design space; shaded interactions are those implemented in the prototype.}
\label{tab:utility}
\resizebox{0.6\columnwidth}{!}{%
\begin{tabular}{ll}
\toprule
\textbf{Visualization Task}      & \hspace{0.25cm}\textbf{Interaction Pairing }                                                                \\
\midrule
Mathematical Operation & \noshade{Hover to add / subtract}                                                             \\
\myhline
Scale Alteration        & \noshade{Pinch on edge to rescale}                                                            \\
\myhline
Encode                  & \begin{tabular}[c]{@{}l@{}}\myshade{Tap to recolor}\\ \noshade{Rotate to switch vis types}\end{tabular} \\
\myhline
Reconfigure          & \begin{tabular}[c]{@{}l@{}} \noshade{Press to flatten}\\ \myshade{Neighbor, stack, assemble to combine} \end{tabular}    \\
\myhline
Abstract / Elaborate & \begin{tabular}[c]{@{}l@{}}\noshade{Double tap to show overview / detail}\\ \noshade{Disassemble to chop}\end{tabular} \\
\myhline
Filter                  & \begin{tabular}[c]{@{}l@{}}\noshade{Swipe to adjust range}\\ \myshade{Cover to hide}\end{tabular}       \\
\myhline
Explore                 & \noshade{Pinch on surface to zoom}                                                            \\
\myhline
Process Control         & \begin{tabular}[c]{@{}l@{}} \noshade{Pick up to initiate}\\ \myshade{Shake to reset}\end{tabular}       \\
\bottomrule
\end{tabular}%
}
\end{table}